\documentclass[conference]{IEEEtran}
\IEEEoverridecommandlockouts
\usepackage{cite}
\usepackage{amsmath,amssymb,amsfonts}
\allowdisplaybreaks[4]
\usepackage{graphicx}
\usepackage{textcomp}
\usepackage{xcolor}
\def\BibTeX{{\rm B\kern-.05em{\sc i\kern-.025em b}\kern-.08em
    T\kern-.1667em\lower.7ex\hbox{E}\kern-.125emX}}
\usepackage{float}
\usepackage{subfig}
\usepackage{booktabs}
\usepackage{tabularx}
\usepackage{ragged2e} 
\usepackage{booktabs} 
\newcolumntype{L}{>{\RaggedRight\hangafter=1\hangindent=0em}X}
\usepackage{bm}
\usepackage{bbm}

\usepackage{amssymb}
\makeatletter

\newcommand{\Rmnum}[1]{\expandafter\@slowromancap\romannumeral #1@}
\makeatother

\usepackage{gensymb}
\definecolor{qq}{HTML}{4c8dae}
\definecolor{qb}{HTML}{48c0a3}
\definecolor{qr}{HTML}{bc7c7c}

\usepackage{stfloats}

\usepackage{multirow}

\usepackage{algorithm}
\usepackage{algpseudocode}

\usepackage{enumitem}
\usepackage{url}
\usepackage{multicol,lipsum}

\usepackage{hyperref}
\hypersetup{
  colorlinks=true,
  linkcolor=blue,
  filecolor=blue,
  urlcolor=blue,
  citecolor=blue
}

\begin{document}

\title{Optical Wireless Ether: Enabling Controlled Dynamic Signal Propagation in OWC Systems}

\author{\IEEEauthorblockN{Hongwei Cui, Soung Chang Liew}

\IEEEauthorblockA{Department of Information Engineering, The Chinese University of Hong Kong, Hong Kong SAR, China}  Email: \{ch021, soung\}@ie.cuhk.edu.hk}

\maketitle

\begin{abstract}
Optical wireless communication (OWC) leverages the terahertz-scale optical spectrum to enable ultra-fast data transfer, offering a compelling alternative to often-congested radio frequency systems. However, the highly directional nature of optical signals and their susceptibility to obstruction inherently limit coverage and reliability, particularly in dynamic indoor environments.
To overcome these limitations, we propose \textit{optical wireless ether (OWE)}, a novel framework that transforms indoor spaces into a dynamically controllable optical propagation medium. OWE employs a distributed network of \textit{ether amplifiers (EAs)}, which act as optical amplifiers with programmable gain values to extend coverage through diffuse reflections while compensating for signal attenuation.
A key challenge in OWE is preventing amplifier saturation from feedback loops. We rigorously derive stability constraints to guarantee system robustness.
Beyond coverage extension, OWE dynamically adjusts EA gains in response to user locations and channel conditions, enhancing signal-to-noise ratio, balancing resource allocation, and suppressing interference.
As the first framework to harness diffuse reflection for controllable optical propagation, we validate OWE’s effectiveness through analytical modeling, simulations, and prototyping. Our work lays the foundation for robust, high-speed indoor OWC networks.
\end{abstract}

\section{Introduction}
Optical wireless communication (OWC) is a promising technology that utilizes light to transmit data, offering significantly higher bandwidth than traditional radio frequency (RF)-based systems. The terahertz-scale optical spectrum enables ultra-fast data transfer rates, making OWC ideal for bandwidth-intensive and latency-sensitive applications such as video streaming, augmented reality (AR), virtual reality (VR), and cloud gaming. Moreover, the license-free nature of optical bands alleviates the congestion plaguing RF spectrum due to the exponential growth of connected devices. These advantages position OWC as a strong candidate for next-generation wireless networks.
\par
However, the inherent directionality of optical signals and their susceptibility to obstruction by opaque objects -- due to shorter electromagnetic wavelengths -- limit coverage and impact connectivity reliability, especially in dynamic indoor environments, where human movement and other transient obstacles can disrupt connectivity.
\par
To overcome this limitation, we propose \textbf{optical wireless ether (OWE)}, a novel framework that transforms indoor spaces into a controlled dynamic optical propagation medium. By adaptively manipulating light propagation, OWE effectively extends signal coverage while enhancing reliability and mobility support.
\begin{figure}
\centering
\includegraphics[width=0.33\textwidth]{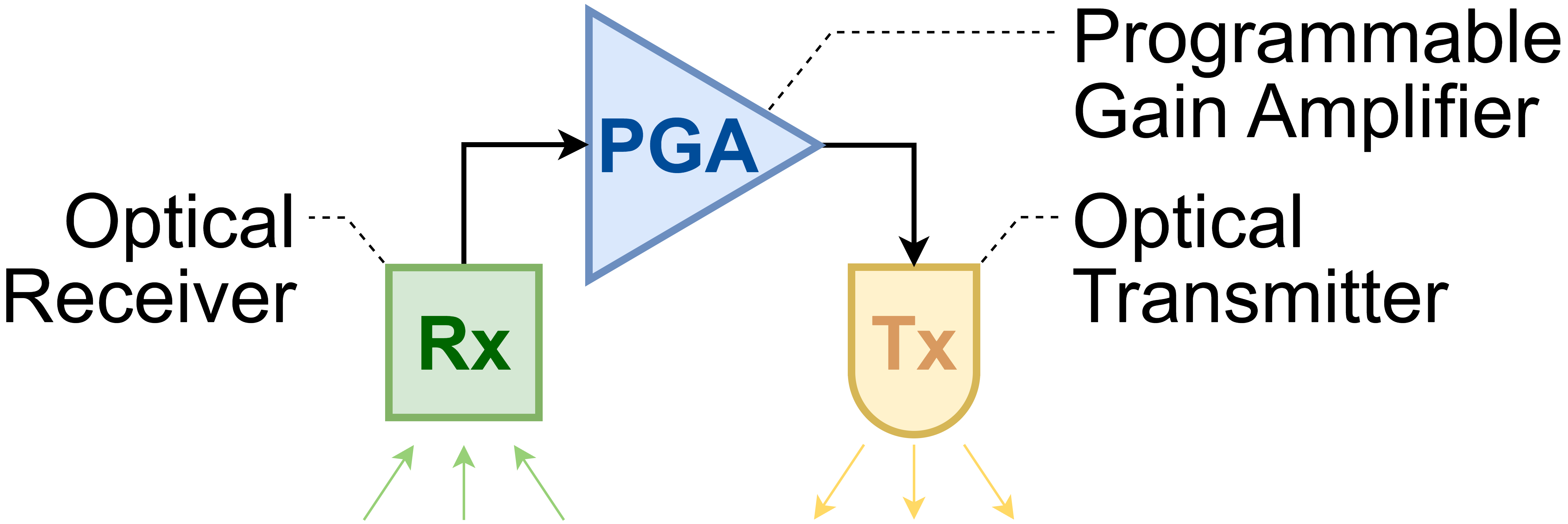}
\caption{Block diagram of the EA.}
\label{EA_Diagram}
\vspace{-0.5cm}
\end{figure}
\par
The OWE framework comprises a distributed network of \textbf{ether amplifiers (EAs)} strategically deployed on ceilings. As illustrated in Fig. \ref{EA_Diagram}, each EA functions by amplifying and retransmitting optical signals at the analog level, without digital processing. This analog operation is critical, as it ensures minimal latency and high-speed propagation through the OWE. Signals within the OWE propagate through diffuse reflections, bouncing between environmental surfaces (e.g., floors, furniture) and the EAs. Each time a signal is received by an EA, it is amplified and re-emitted, effectively extending coverage while compensating for attenuation.
\par
A key innovation of OWE lies in its ability to dynamically optimize the signal propagation by adjusting the programmable gain amplifiers (PGAs) embedded in each EA. By coordinating the gains across multiple EAs, we can precisely modify the OWE channel characteristics. This dynamic control allows us to redirect light to circumvent obstacles and specifically enhance signal quality based on station locations and channel conditions. 
Beyond extending signal coverage, OWE offers three key benefits:
1) SNR Optimization: Maximizes the signal-to-noise ratio (SNR) within individual basic service sets (BSSs) \cite{gast2005802}, improving communication quality between access points (APs) and stations. 
2) Fair Resource Allocation: Enables balanced distribution of communication resources across multiple BSSs, maintaining consistent signal quality for concurrent signal streams. 
3) Interference Mitigation: Reduces mutual interference when multiple APs serve different stations simultaneously.
\par
A critical challenge for the OWE design arises from the closed-loop interactions among EAs: Each EA’s input may aggregate signals from multiple neighboring EAs, while its output can feed back into the same set of EAs. Without careful control, this can lead to amplifier saturation, signal distortion, and network instability. Thus, the critical design challenge is to dynamically tune EA gains to prevent saturation while optimizing coverage and signal quality. 
This paper derives the fundamental constraints for saturation prevention and demonstrates how to optimize system performance without violating these constraints.
\par
Our overall contributions are as follows:
1) Proposal of OWE: To our knowledge, OWE is the first framework that exploits diffuse reflection to create a controllable, dynamic optical signal propagation medium, effectively overcoming conventional OWC limitations.
2) Design and Implementation of EAs: We provide the design of EAs, which perform analog amplification and retransmission of optical signals to enable low-latency, high-speed propagation within the OWE framework.
3) Ensuring OWE Stability: We identify and address the critical challenge of maintaining network stability and preventing amplifier saturation caused by closed-loop EA interactions.
4) Dynamic Channel Optimization: We show how to coordinate EA gains dynamically to enhance OWE's performance across key metrics -- improving SNR in single-BSS scenarios; and enhancing SINR, resource allocation fairness, and mutual interference suppression in multi-BSS scenarios.
\par
We envision OWE as a foundational framework for future advancements in optical wireless communication, paving the way for reliable, high-speed indoor connectivity.
\section{System Model}
\subsection{System Design Overview}
\begin{figure}
\centering
\includegraphics[width=0.5\textwidth]{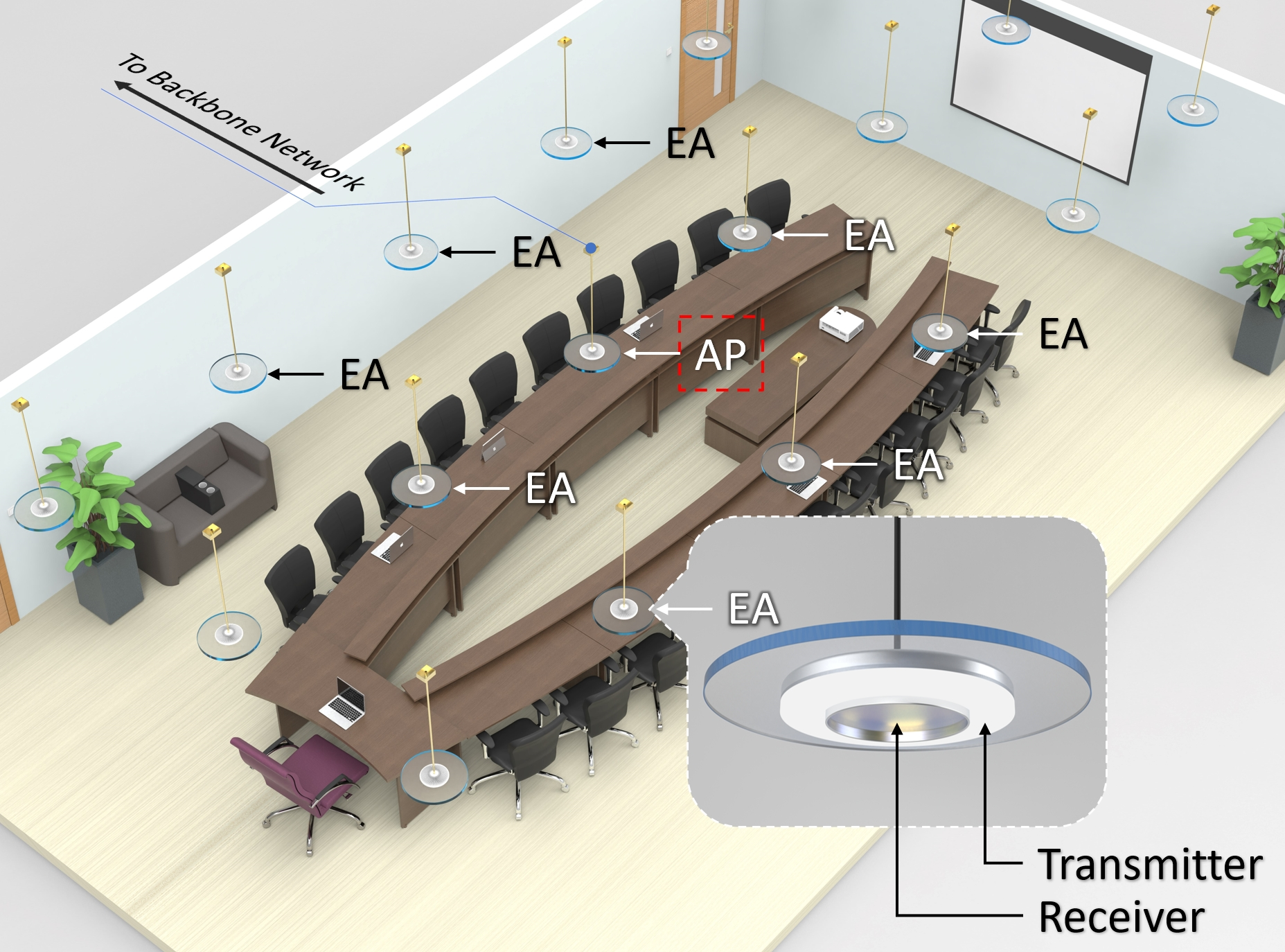}
\caption{Meeting room application scenario of OWE. An AP and multiple EAs are mounted on the ceiling.}
\label{meeting_room}
\end{figure}

OWE comprises a distributed network of EAs that dynamically extend coverage by amplifying and retransmitting optical signals in real time.
In our system model, both APs and EAs serve dual roles: they function as light sources for illumination while simultaneously enabling communication. These devices are equipped with downward-facing optical transmitters and receivers and typically ceiling-mounted in a uniform grid (though non-uniform deployments are also feasible) to ensure seamless coverage for both stationary and mobile user stations, as depicted in Fig. \ref{meeting_room}.
Furthermore, OWE uses intensity modulation for communication. Flickering is not an issue because the modulation frequencies operate in the megahertz (MHz) range, which is imperceptible to the human eye \cite{wikipediacontributors_2019_flicker}.
\par
\begin{figure}[htbp]
\centering
\includegraphics[width=0.45\textwidth]{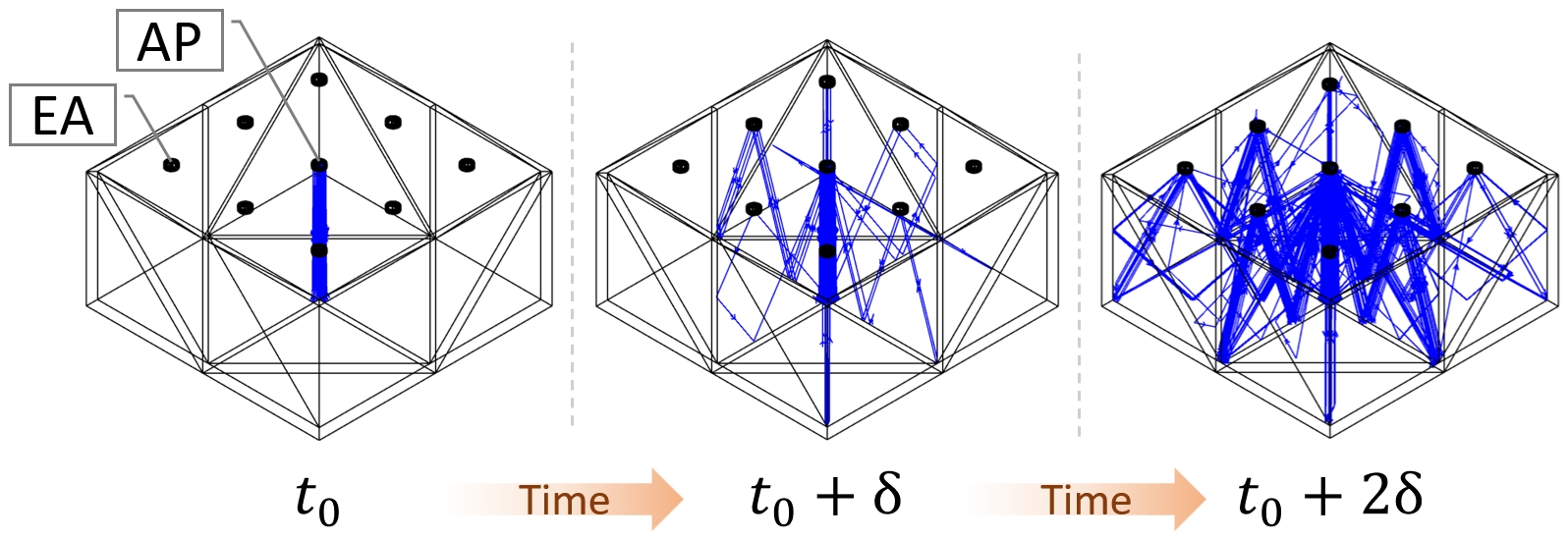}
\caption{Signal propagation in OWE via diffuse reflections.}
\label{reflecting_path}
\vspace{-0.5cm}
\end{figure}

OWE primarily relies on \textbf{diffuse reflections} from surfaces such as floors for signal propagation (Fig. \ref{reflecting_path}). Unlike digital regenerators -- which decode, reconstruct, and retransmit signals -- each EA functions as an analog repeater, amplifying and retransmitting incoming signals without modification.
This approach has three key advantages:
1) Hardware Simplicity: EAs are simple devices that require no complex circuitry, backbone network connectivity, or dedicated transceivers for inter-EA communication, significantly reducing system complexity.
2) Near-Zero Latency: By leveraging analog amplification, EAs bypass the delays inherent in digital relays, such as digital signal processing at the PHY layer and MAC-layer coordination/retransmission, making them suited for latency-sensitive applications.
3) Mobility \& Resilience: Diffuse reflection inherently enhances mobility support and improves resilience against blockages compared to line-of-sight (LOS) channels.
\par
A limitation of this analog repeater implementation is that EAs amplify not only the desired signal but also any accumulated noise indiscriminately. Unlike digital regenerators that can filter or reconstruct signals to mitigate noise, EAs propagate noise alongside the signal. Furthermore, the closed-loop interactions among EAs can lead to positive feedback and saturate amplifiers. To address these challenges, we developed a detailed \textbf{system model}, with \textbf{noise modeling} to quantify noise accumulation, and analyzed system \textbf{stability conditions} to prevent amplifier saturation.

\subsection{Diffuse Reflection Channel Model Between EAs}
The diffuse reflection channels between EAs form the foundational propagation mechanism of the OWE system, determining its overall signal propagation characteristics. We first establish a comprehensive model for the inter-EA diffuse reflection channel.
\par
The optical transmitter in each EA is modeled as a Lambertian emitter \cite{ghassemlooy2019optical} with a radiant intensity $S\left( \phi  \right)$, given by
\begin{equation}
\label{lambertian_emission}
S(\phi ) = {{P_t}\frac{{m + 1}}{{2\pi }}}{\cos ^m}\phi \text{ ,}
\end{equation}
where $P_t$ is the transmitting optical power; $\phi$ is the angle between the emitter’s normal and the radiated light beam; $m =  - \frac{{\ln 2}}{{\ln \left( {\cos {\Phi _{1/2}}} \right)}}$ is the Lambertian order, and ${\Phi _{1/2}}$ is the half-power angle.
\par
\begin{figure}
\centering
\includegraphics[width=0.4\textwidth]{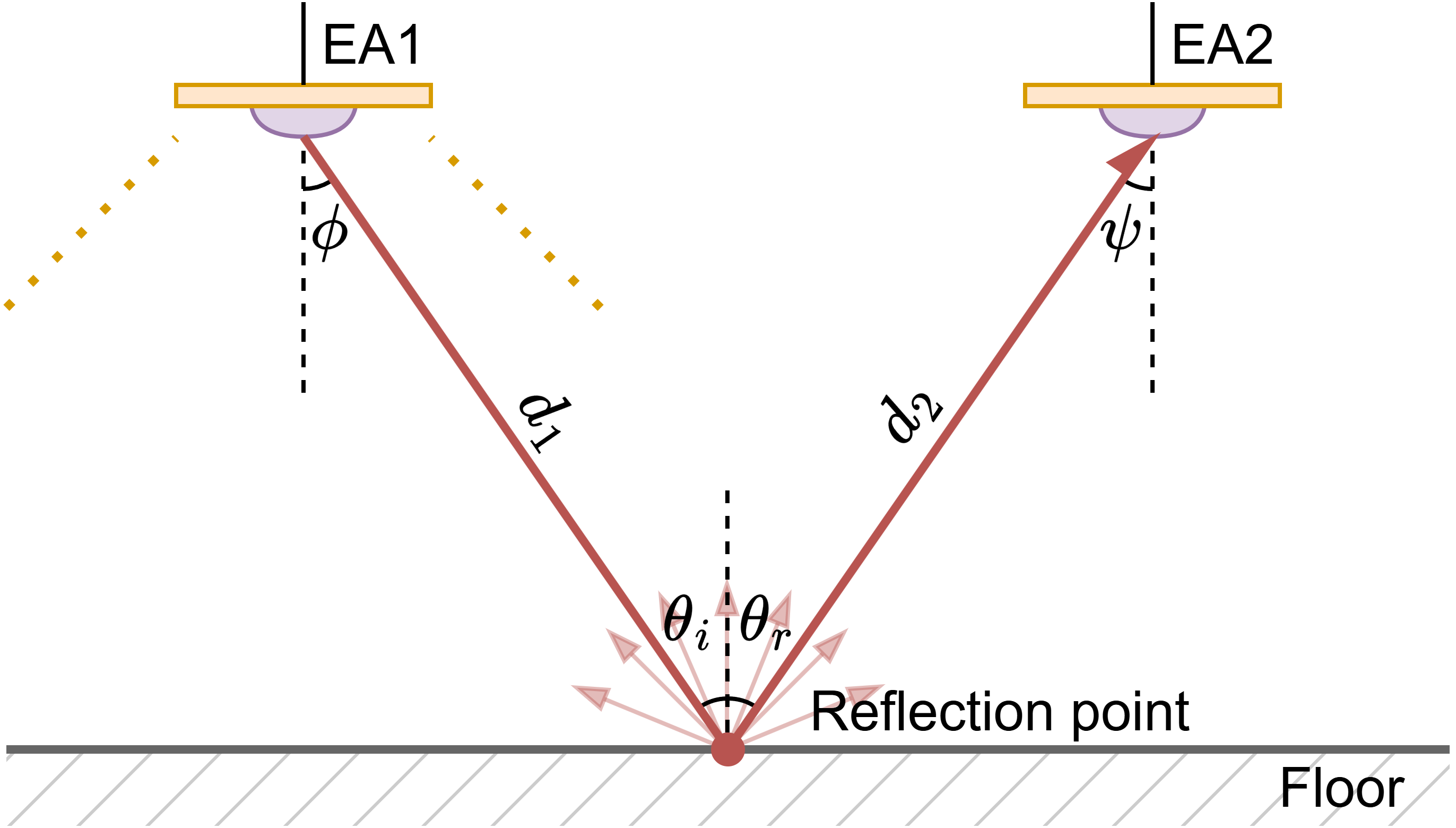}
\caption{Signal propagation between EAs via floor diffuse reflections.}
\label{diffuse_reflection}
\vspace{-0.5cm}
\end{figure}
The floor is modeled with reflectivity $\rho$ and Lambertian-like scattering (diffuse reflection), as shown in Fig. \ref{diffuse_reflection}. The reflected power reaching the receiver per unit floor area is
\begin{equation}
\label{lambertian_reflection}
d{P_r} = S\left( \phi  \right) \cdot \frac{{\rho \cos {\theta _i}\cos {\theta _r}}}{{\pi d_1^2d_2^2}} \cdot {A_{\text{eff}}} \cdot d{A_{\text{floor}}} \text{ .}    
\end{equation}
The symbols in Eqn. (\ref{lambertian_reflection}) above are defined as follows: $d_1$ denotes the distance from the EA transmitter to the reflection point, and $d_2$ the distance from the reflection point to the EA receiver. The angle of incidence at the floor is $\theta_i$, while $\theta_r$ represents the angle of reflection toward the receiver. The effective receiver area, $A_{\text{eff}}$, corresponds to the projected area of the photodetector and is given by ${A_{\text{eff}}} = {A_r}\cos \psi$, where $A_r$ is the physical receiver area, and $\psi$ is the angle between the receiver’s normal and the incoming ray.
\par
Although an ideal large-area photodiode (PD) is theoretically suitable for our system, practical implementation faces challenges such as high manufacturing costs and limited bandwidth due to increased junction capacitance \cite{ghassemlooy2019optical}. To address these issues, a cost-effective solution is to integrate a non-imaging concentrator -- specifically, a compound parabolic concentrator -- before the PD. The concentrator enhances the PD’s effective collection area through a geometric gain factor $g_C$, defined as
\begin{equation}
\label{concentrator_gain}
{g_C}\left( \psi  \right) = \frac{{{\tau ^2}}}{{{{\sin }^2}\left( {{\Psi _C}} \right)}} \cdot {\rm{rect}}\left( {\frac{\psi }{{{\Psi _C}}}} \right) \text{ ,}   
\end{equation}
where ${\rm{rect}}(\cdot)$ denotes the rectangular function (equal to 1 for ${\psi  \le {\Psi _C}}$ and 0 otherwise), $\tau$ represents the refractive index of the concentrator material, and ${\Psi _C}$ is the half-FOV (acceptance angle) of the receiver with concentrator.
\par
Additionally, the optical band-pass filter at the receiver can be represented as ${T_s}\left( \psi  \right)$. Assuming OWE operates across the full spectrum of the transmission signal, the filter can be omitted in the hardware implementation and set ${T_s}\left( \psi  \right) = 1$.
\par
The total optical power hitting the receiver detector is obtained by integrating over the reflection area, as given in Eqn. (\ref{channel_model}).
\begin{table*}[hbp]
\centering
\begin{equation}
\label{channel_model}
{P_r} = 
\int \left[ {{g_C}\left( \psi  \right) \cdot {T_s}\left( \psi  \right)} \right] d{P_r}
= \int \left[{\left( {{P_t}\frac{{\left( {m + 1} \right)}}{{2\pi }}{{\cos }^m}\phi } \right) \cdot \left( {\frac{{\rho \cos {\theta _i}\cos {\theta _r}}}{{\pi d_1^2d_2^2}} \cdot {A_r}\cos \psi } \right)} \right. \left. \cdot \left( {\frac{{{\tau^2}}}{{{{\sin }^2}\left( {{\Psi _C}} \right)}} \cdot {\rm{rect}}\left( {\frac{\psi }{{{\Psi _C}}}} \right) \cdot {T_s}\left( \psi  \right)} \right) \right] d{A_{{\rm{floor}}}}
\end{equation}
\end{table*}
\subsection{EA Noise Model}
\begin{figure}
\centering
\includegraphics[width=0.40\textwidth]{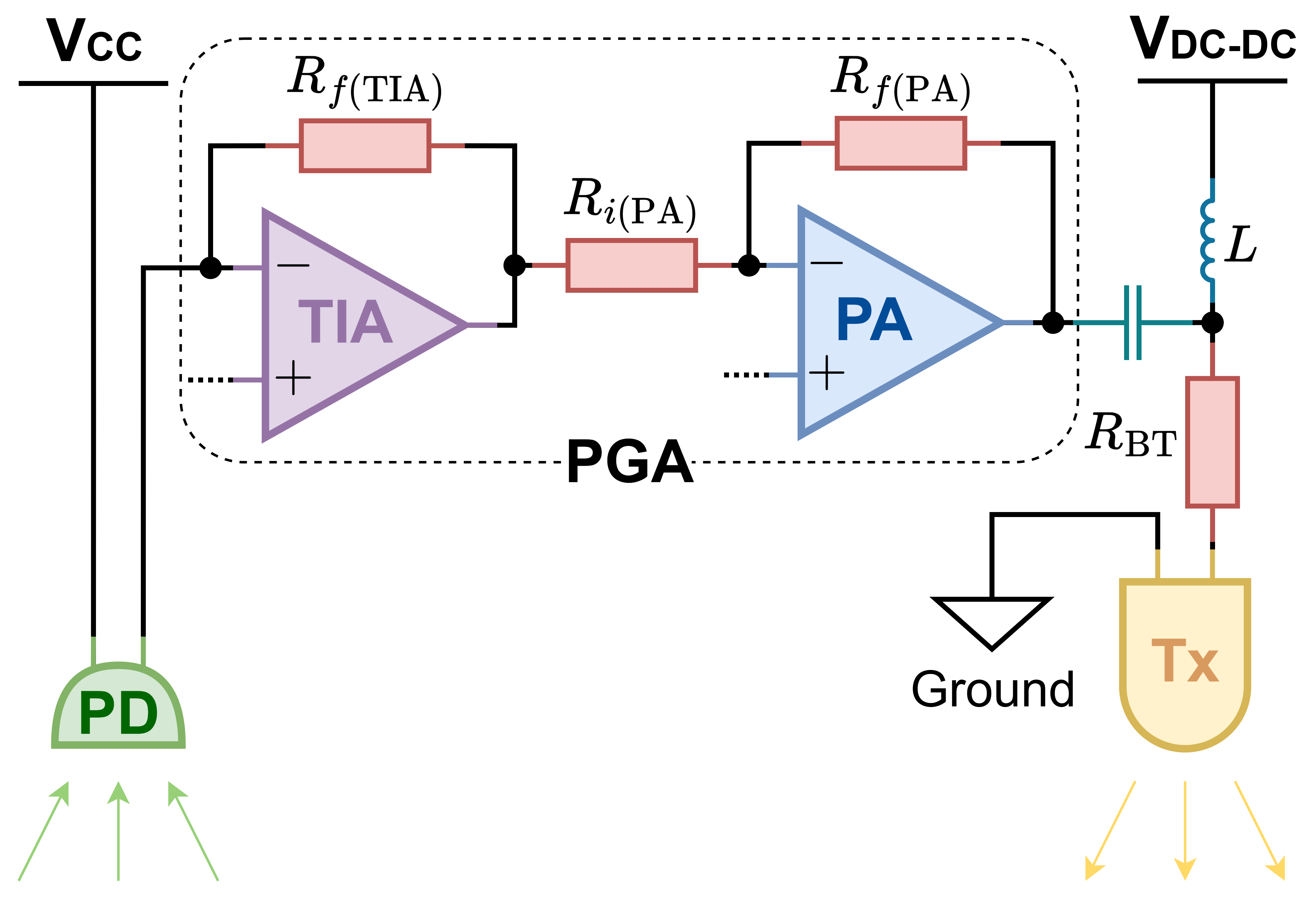}
\caption{Circuit model for the EA. The PGA in the EA is constructed using a cascaded TIA and a PA.}
\label{ea_circuit_model}
\vspace{-0.5cm}
\end{figure}
EA is an active electrical device and inevitably introduces noise into the signal it amplifies. The primary noise sources in the EA include the photodiode, transimpedance amplifier (TIA), power amplifier (PA), and the switching DC-DC converter, as illustrated in Fig. \ref{ea_circuit_model}. To minimize noise in sensitive amplification stages, low-noise linear regulators (LNRs) are used to power the amplifiers, ensuring an ultra-clean supply and negligible power-related noise. However, the LED bias current -- typically large to achieve sufficient optical output -- requires a switching DC-DC converter for higher efficiency and driving capability.
\par
A detailed noise analysis is provided in Appendix A. The noise component in the \textbf{current} going through the LED to modulate the radiated light intensity is given by
\begin{equation}
\label{LED_current_noise}
{n_{\text{LED}}} = \frac{{{G_{\text{PA}}}{Z_{\text{TIA}}}}}{{{R_{\text{BT}}}}}\left( {{n_{\text{PD}}} + {n_{\text{TIA}}} + {n_{\text{PA,TIA}}}} \right) + {n_{\text{DC-DC}}} \text{ ,}
\end{equation}
where $n_{\text{PD}}$ is the PD’s current noise, $n_{\text{TIA}}$ is the TIA’s input-referred current noise, $n_{\text{PA,TIA}}$ is the PA’s current noise referred to the TIA input (for a unified model), and $n_{\text{DC-DC}}$ is the DC-DC converter’s current noise.
As depicted in Fig. \ref{ea_circuit_model}, the PA drives the LED through a series resistor $R_{\text{BT}}$, which stabilizes the current and prevents excessive loading. Given the LED’s negligible small-signal impedance at its DC operating point, $R_{\text{BT}}$ dominates the PA's load. The PA's voltage gain is ${G_{\text{PA}}} =  - \frac{{{R_{f\left( {\text{PA}} \right)}}}}{{{R_{i\left( {\text{PA}} \right)}}}}$, while the TIA’s transimpedance is ${Z_{\text{TIA}}} = {R_{f\left( {\text{TIA}} \right)}}$ for simple resistive feedback.
\par
Notably, the EA-induced noise is modeled as two separate components for accuracy:
1) Gain-dependent noise: The PD, TIA, and PA introduce noise that is amplified along with the signal, making these noise components dependent on the EA's gain settings.
2) Additive noise: The DC-DC converter's noise (in the LED bias current) manifests as purely additive noise at the output.

\subsection{Overall OWE Channel Model with EA Effects}
\begin{figure}
\centering
\includegraphics[width=0.48\textwidth]{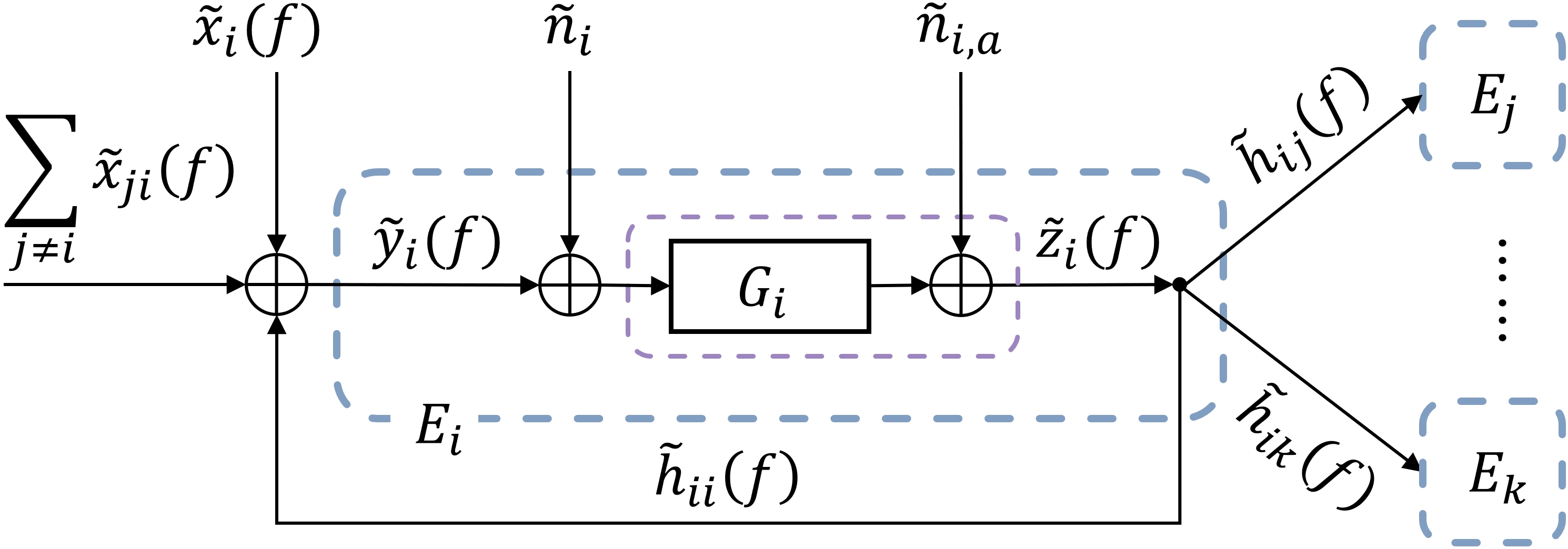}
\caption{The EA's signal flow diagram.}
\label{system_diagram}
\end{figure}
The gain configurations of the EAs determine the channel characteristics of the constructed OWE. Without loss of generality, Fig. \ref{system_diagram} illustrates the signal flow through the \textit{i}-th EA, which is mathematically described by
\begin{equation}
\label{signal_flow}
\left\{  \begin{aligned}
&{\tilde z_i}(f) = {G_i}\left( {{{\tilde y}_i}(f) + {{\tilde n}_i}} \right) + {\tilde n_{i,a}}\\
&{{\tilde y}_i}(f) = {{\tilde x}_i}(f) + \sum\limits_{j \ne i} {{{\tilde x}_{ji}}(f)}  + {{\tilde h}_{ii}}(f){{\tilde z}_i}(f)
\end{aligned} \right. \text{\ ,}
\end{equation}
where all variables are frequency-domain representations (denoted by $\tilde{\cdot}\ $) of their respective signals. The definitions of symbols are provided in Table \ref{Table1}.
\begin{table}
\centering
\begin{tabularx}{0.5\textwidth}{l L}
\toprule
\textbf{Symbols}&\textbf{Definitions}\\  \midrule 
$E_i$&The \textit{i}-th EA in the OWE system.\\ [0.0cm]
$G_i$&Gain of the \textit{i}-th EA, given by $G_i=\frac{{{G_{\text{PA}}}{Z_{\text{TIA}}}}}{{{R_{\text{BT}}}}}$.\\ [0.0cm]
$\tilde{h}_{ii},\tilde{h}_{ij}$&Channel responses in the OWE system accounting for both the transmitter-to-receiver optical path and PD/LED optical-current conversion characteristics, where:
\begin{itemize}
\setlength{\itemindent}{-1em}
\item $\tilde{h}_{ii}$ is the \textit{self}-channel response of the \textit{i}-th EA.
\item $\tilde{h}_{ij}$ is the \textit{cross}-channel response from EA \textit{i} to EA \textit{j}.
\end{itemize} \\ [-0.2cm]

$\tilde{x}_i,\tilde{x}_{ji}$&Photocurrent outputs at the \textit{i}-th EA's PD, where:
\begin{itemize}
\setlength{\itemindent}{-1em}
\item $\tilde{x}_i$ results from optical power transmitted by the station.
\item $\tilde{x}_{ji}$ results from optical power transmitted by the \textit{j}-th EA.
\end{itemize} \\ [-0.2cm]

$\tilde{y}_i$&Noiseless aggregate photocurrent output at the \textit{i}-th EA's PD.\\ [0.0cm]
$\tilde{z}_i$&LED drive current for the \textit{i}-th EA.\\ [0.0cm]

$\tilde{n}_i,\tilde{n}_{i,a}$&Noise components in the \textit{i}-th EA, where:
\begin{itemize}
\setlength{\itemindent}{-1em}
\item $\tilde{n}_i$: Gain-dependent noise, $\tilde{n}_i={{\tilde{n}_{\text{PD}}} + {\tilde{n}_{\text{TIA}}} + {\tilde{n}_{\text{PA,TIA}}}}$.
\item $\tilde{n}_{i,a}$: Additive noise, $\tilde{n}_{i,a}={\tilde{n}_{\text{DC-DC}}}$.
\end{itemize} \\ [-0.2cm]

$N$ & Number of EAs in the OWE system.\\
\bottomrule
\end{tabularx}
\caption{Definitions of symbols in Fig. \ref{system_diagram}.}
\label{Table1}
\vspace{-0.5cm}
\end{table}
Combining the two equations in (\ref{signal_flow}), we can obtain 
\begin{equation}
\label{sys_equ}
\resizebox{.9\hsize}{!}{$
\begin{aligned}
{{\tilde y}_i}(f)
={{\tilde x}_i}(f) &+ \left[ {\sum\limits_{j \ne i} {\left( {{{\tilde h}_{ji}}(f){G_j}{{\tilde y}_j}(f)} \right)}  + {{\tilde h}_{ii}}(f){G_i}{{\tilde y}_i}(f)} \right] \\
&+ \left( {\sum\limits_{j \ne i} {{{\tilde h}_{ji}}(f){G_j}{{\tilde n}_j}}  + {{\tilde h}_{ii}}(f){G_i}{{\tilde n}_i}} \right) \\ 
&+ \left( {\sum\limits_{j \ne i} {{{\tilde h}_{ji}}(f){{\tilde n}_{j,a}}}  + {{\tilde h}_{ii}}(f){{\tilde n}_{i,a}}} \right) \text{ .}
\end{aligned}
$}
\end{equation}
The left-hand side of (\ref{sys_equ}) represents the combined received signal at the \textit{i}-th EA, ${\tilde y}_i(f)$. The right-hand side captures the signal input from the station, as well as the influence of other EAs and noise components.
\par \noindent
Let us define 
\begin{equation}
    \label{symbol_definitions}
    \begin{aligned}
        &\mathbf{G} = {\mathop{\rm Diag}\nolimits} \left( {{G_1},{G_2}, \ldots ,{G_N}} \right), \quad
        \mathbf{H}\left( {i,j} \right) = {{\tilde h}_{ij}}(f),\\
        &\mathbf{\tilde x} = {\left( {{\tilde{x}_1},{\tilde{x}_2}, \ldots ,{\tilde{x}_N}} \right)^T}, \quad
        \mathbf{\tilde y} = {\left( {{\tilde{y}_1},{\tilde{y}_2}, \ldots ,{\tilde{y}_N}} \right)^T},\\
        &\mathbf{\tilde n} = {\left( {{\tilde{n}_1},{\tilde{n}_2}, \ldots ,{\tilde{n}_N}} \right)^T}, \quad {{\mathbf{\tilde n}}_\mathbf{a}} = {\left( {{\tilde{n}_{1,a}}, {\tilde{n}_{2,a}}, \ldots {\tilde{n}_{N,a}}} \right)^T},
    \end{aligned}
\end{equation}
where $N$ is the number of EAs. Eqn. (\ref{sys_equ}) can be generalized to all $N$ EAs, giving $N$ equations. These equations can be collected together in a succinct matrix form:
\begin{equation}
\label{mat_sys_equ}
\begin{array}{c}
\left( {\mathbf{I} - {\mathbf{H}^T}\mathbf{G}} \right)\mathbf{\tilde y} = \mathbf{\tilde x + }{\mathbf{H}^T}\mathbf{G\tilde n + }{\mathbf{H}^T}{{\mathbf{\tilde n}}_\mathbf{a}} \text{ .}
\end{array}
\end{equation}
Equation (\ref{mat_sys_equ}) describes the relationship between the external signal inputs of OWE from the station, $\mathbf{\tilde{x}}$, and the combined received signals at EAs, $\mathbf{\tilde{y}}$. Note that $\mathbf{\tilde{y}}$ can be interpreted as a \textbf{channel response} to the input $\mathbf{\tilde{x}}$. Therefore, by adjusting the gain values of EAs, $\mathbf{G}$, we can change the light propagation characteristics of the OWE.

\subsection{Key Challenge in OWE Stability}
Increasing the gains of EAs can compensate for the signal loss, thereby extending the signal coverage and improving signal quality.
However, the closed-loop interactions among EAs present a fundamental challenge in OWE implementation. Each EA’s input combines signals from multiple neighboring EAs, while its output simultaneously feeds back into the same network. Improper gain settings among EAs can generate positive feedback loops, leading to amplifier saturation and signal distortion.
\par
Consequently, a core design challenge is to dynamically optimize PGA gains across all EAs -- preventing amplifier saturation while ensuring sufficient signal coverage.
\par
For analytical convenience, we derive a noiseless channel model (\ref{mat_sys_equ_noiseless}) from (\ref{mat_sys_equ}). The resulting gain constraints remain
valid even with noise, i.e., 1) if saturation occurs with signal, even a tiny amount of noise without signal will trigger amplifier saturation; and 2) if no saturation occurs with signal, the inclusion of noise will not cause saturation either. 
\begin{equation}
\label{mat_sys_equ_noiseless}
\begin{aligned}
& \left( {\mathbf{I} - {\mathbf{H}^T}\mathbf{G}} \right)\mathbf{\tilde y} = \mathbf{\tilde x} \\
&{\ \ \ \ \ \ \ \ \ \ \ \ \ \ \ \ \ }\Downarrow \\
\mathbf{\tilde y} &=\left[ {\mathbf{I} + \mathbf{H}^T \mathbf{G + }{\left(\mathbf{H}^T\mathbf{G}\right)^2} +  \cdots } \right]\mathbf{\tilde x}\\
&= \lim_{m\rightarrow \infty}\left[ {\mathbf{I} - {\left(\mathbf{H}^T\mathbf{G}\right)^m}} \right]{\left( {\mathbf{I} - \mathbf{H}^T\mathbf{G}} \right)^{ - 1}}\mathbf{\tilde x}
\end{aligned}
\end{equation}
According to (\ref{mat_sys_equ_noiseless}), the OWE is stable if and only if the EAs' gain settings fulfill
\begin{equation}
\label{converge_cond}
{\left({\mathbf{H}^T}\mathbf{G}\right)^m} \to 0\textit{, as }m \to \infty \text{ .}
\end{equation}
The matrix ${\mathbf{H}^T}\mathbf{G}$ is square. If it adopts an eigen decomposition and satisfies condition (\ref{converge_cond}), the absolute value of each eigenvalue must fulfill $\left| {{\lambda_{\mathbf{H}^T\mathbf{G}}}} \right| < 1$.  Equivalently, the \textbf{stability condition} can be stated as:
\par
\textbf{The spectral radius of $\mathbf{H}^T\mathbf{G}$ must be less than $1$.}
\par
Further, since $\left| {{\lambda_{\mathbf{H}^T\mathbf{G}}}} \right| < 1$, we have $\det {\left( {\mathbf{I} - {\mathbf{H}^T}\mathbf{G}} \right)}>0$, ensuring that ${\left( {\mathbf{I} - {\mathbf{H}^T}\mathbf{G}} \right)}$ is invertible under the OWE stability constraint. 
Let $\mathbf{A} = {\left( {\mathbf{I} - {\mathbf{H}^T}\mathbf{G}} \right)^{ - 1}}$, we can then rewrite (\ref{mat_sys_equ}) as
\begin{equation}
\label{eq10}
\mathbf{\tilde y} = \mathbf{A}\left[ {\mathbf{\tilde x + }{\mathbf{H}^T}\left( {\mathbf{G\tilde n + }{{\mathbf{\tilde n}}_\mathbf{a}}} \right)} \right] \text{ .}
\end{equation}
\subsection{SNR of a Signal Propagating Through OWE}
\begin{figure}
\centering
\includegraphics[width=0.48\textwidth]{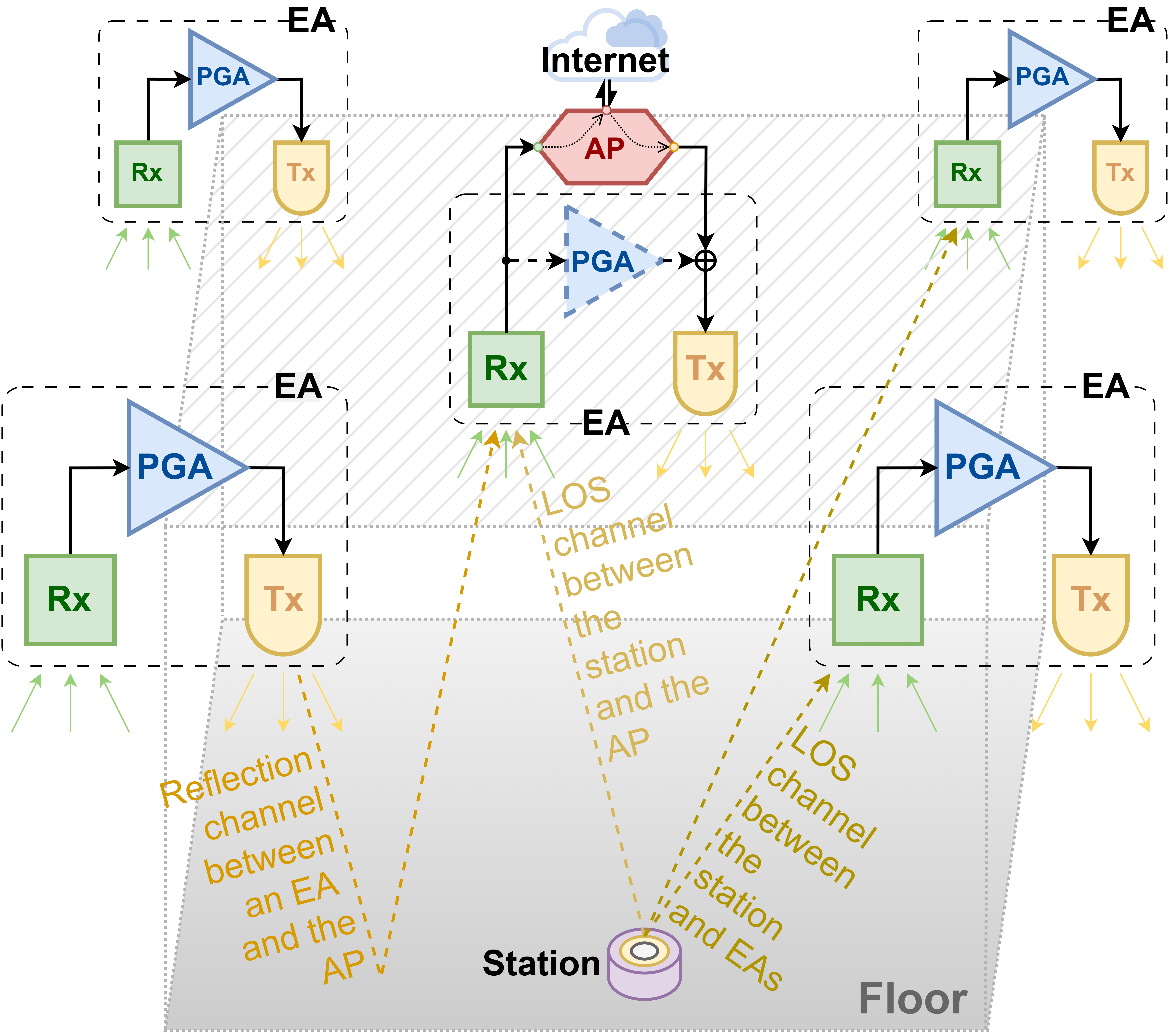}
\caption{Signal propagations from the OWE to the AP. For simplicity, the figure shows the reflection path from only one EA to the AP.}
\label{ea_ap}
\vspace{-0.5cm}
\end{figure}

Figure \ref{ea_ap} shows the two main uplink signal propagation paths in the OWE:
1) Line-of-Sight Paths: The station's signal reaches the AP directly when located within its coverage area.
2) Non-Line-of-Sight Paths: The signal reaches the AP through diffuse reflection and amplification in the OWE.
Accounting for both path types, the received signal at the AP can be expressed as
\begin{equation}
\label{eq11}
\begin{aligned}
{{\tilde r}_{A}} 
&={{\tilde h}_{\text{SA}}}\left( f \right){{\tilde s}_{\text{S}}}\left( f \right)
+ {{\tilde n}_{\text{A}}} + {\left( {\mathbf{G}{{\mathbf{\tilde h}}_{\text{EA}}}} \right)^T} \mathbf{\tilde y}\\
&={{\tilde h}_{\text{SA}}}\left( f \right){{\tilde s}_{\text{S}}}\left( f \right)
+ {{\tilde n}_{\text{A}}} \\
&\quad + {\left( {\mathbf{G}{{\mathbf{\tilde h}}_{\text{EA}}}} \right)^T}\mathbf{A}\left[ {{{\mathbf{\tilde h}}_{\text{SE}}}{{\tilde s}_{\text{S}}}\left( f \right) + {\mathbf{H}^T}\left( {\mathbf{G\tilde n + }{{\mathbf{\tilde n}}_\mathbf{a}}} \right)} \right] \text{ .}
\end{aligned}
\end{equation}
In (\ref{eq11}), ${{\tilde r}_{A}}$ denotes the received signal at the AP, ${{\tilde n}_{\text{A}}}$ denotes the AP's noise. ${{\tilde s}_{\text{S}}}$ is the station's transmitted signal, and ${{\tilde h}}_{\text{SA}}$ denotes the LOS channel response from the station to the AP (typically zero when no direct path exists). The channel responses are vector quantities:
{\footnotesize
${{\mathbf{\tilde h}}_{\text{EA}}} = {\left( {{{\tilde h}_{\text{1,A}}}\left( f \right),{{\tilde h}_{\text{2,A}}}\left( f \right), \ldots ,{{\tilde h}_{\text{N,A}}}\left( f \right)} \right)^T}$}
describes the channel from the EAs to the AP, and
{\footnotesize
${{\mathbf{\tilde h}}_{\text{SE}}} = {\left( {{{\tilde h}_{\text{S,1}}}\left( f \right),{{\tilde h}_{\text{S,2}}}\left( f \right), \ldots ,{{\tilde h}_{\text{S,N}}}\left( f \right)} \right)^T}$}
represents the channel from the station to the EAs, where ${{\mathbf{\tilde h}}_{\text{SE}}}{{\tilde s}_{\text{S}}}\left( f \right)$ corresponds to $\mathbf{\tilde x}$ in (\ref{eq10}).
\par
According to (\ref{eq11}), the SNR of the signal received by the AP from the station is
\begin{equation}
\label{ap_snr}
\text{SNR}_{\text{SA}} = \frac{{{{\left| {{{\tilde h}_{\text{SA}}}\left( f \right) + {{\left( {\mathbf{G}{{\mathbf{\tilde h}}_{\text{EA}}}} \right)}^T}\mathbf{A}{{\mathbf{\tilde h}}_{\text{SE}}}} \right|}^2}{{\left| {{{\tilde s}_{\text{S}}}\left( f \right)} \right|}^2}}}{{{{\left| {{{\left( {\mathbf{G}{{\mathbf{\tilde h}}_{\text{EA}}}} \right)}^T}\mathbf{A}{\mathbf{H}^T}\left( {\mathbf{G\tilde n + }{{\mathbf{\tilde n}}_\mathbf{a}}} \right) + {{\tilde n}_{\text{A}}}} \right|}^2}}} \text{ .}
\end{equation}
\par
For the reciprocal case, where the station receives signals from the AP, the SNR relationship follows $\text{SNR}_{\text{SA}} \propto \text{SNR}_{\text{AS}}$ due to the aligned propagation paths in both uplink and downlink.
This alignment holds because OWC typically employs intensity modulation/direct detection (IM/DD) \cite{wikipediacontributors_2025_intensity}, where the signal wavelength\footnote{The frequency of intensity modulation is not the same as the optical frequency of the light. The light itself in our system does not cause constructive-destructive interference because the LEDs in our system are not coherent light sources like lasers, and the dimensions of PDs are typically tens of thousands of optical radiation wavelengths \cite{ghassemlooy2019optical, al2018optical}.} (on the order of meters) is far larger than the minor positional deviations (centimeter-scale) between optical transmitters and receivers in the EAs, stations, and APs. As a result, the path differences have a negligible impact on the channel symmetry.

\section{OWE Channel Optimization}
Beyond extending signal coverage while ensuring system stability, the channel characteristics and performance of OWE can be further enhanced by optimizing the gain values of EAs across different scenarios. We examine two key use cases: a single basic service set (BSS) and multiple BSSs.
\subsection{Single-BSS Scenario}
In single-BSS scenarios, OWE facilitates communication between the station and the AP. We aim to optimize the EA gain values to maximize the \textbf{SNR} for the signal stream:
\begin{equation}
\label{eq17}
\begin{aligned}
    &\max_\mathbf{G} \text{SNR}_{\text{SA}} \\
    \textit{s.t.} {\ \ \ \ }
    & \max|\lambda_{{\mathbf{H}^T}\mathbf{G}}|<1,{\ }\\
    & G_i\geq0,{\ }i=1,\dots, N \text{ .}
\end{aligned}
\end{equation}
To avoid potential destructive interference among EAs and simplify the analysis, we disregard the phase shift introduced by the EA and treat $G_i$ as non-negative real-valued quantities. To achieve this, EAs must be designed to provide positive amplification, and any phase shift caused by reactive components (e.g., capacitors and inductors) should be compensated using a properly designed all-pass filter in the hardware implementation \cite{THOMPSON2014531}.
\par
The general representation of $\text{SNR}_{\text{SA}}$ in (\ref{ap_snr}) can be simplified in the following steps to derive (\ref{snr_sa}):
1) Removal of the LOS Component: The LOS component is omitted since it does not affect OWE optimization.
2) Negligible AP Noise: The AP's noise, ${{\tilde n}_{\text{A}}}$, is disregarded, as it is negligible compared to the received signal noise.
3) Direct AP-EA Connection: As shown in Fig. \ref{ea_ap}, the AP connects directly to an EA, bypassing its PGA. Suppose the AP is connected to the \textit{i}-th EA in the OWE, then, the term $\mathbf{G}{{\mathbf{\tilde h}}_{\text{EA}}}$ in (\ref{ap_snr}) reduces to a standard basis vector $\mathbf{w}_{\text{EA}}^T=[0,0,\dots,0, 1, 0\dots, 0]$, where only the \textit{i}-th element is 1.
4) Single-EA Signal Reception: We first consider the simplest case where only the \textit{j}-th EA receives the station's signal. Such EAs are referred to as \textbf{entry EAs}. Let ${{\mathbf{\tilde h}}_{\text{SE}}}{{\tilde s}_{\text{S}}} = \mathbf{w}_{\text{SE}} {\tilde v}$, where $\mathbf{w}_{\text{SE}}^T=[0,0,\dots,0, 1, 0,\dots, 0]$ is a standard basis vector with the \textit{j}-th element set to 1, and ${\tilde v}$ is the photocurrent amplitude at the entry EA.
\begin{align}
\label{snr_sa}
& \max_\mathbf{G} \text{SNR}_{\text{SA}} 
\Leftrightarrow
\max_\mathbf{G} 
\frac{{{{\left|\mathbf{w}_{\text{EA}}^T\mathbf{A}{\left(\mathbf{w}_{\text{SE}} {\tilde v}\right)} \right|}^2}}}{{{{\left| {\mathbf{w}_{\text{EA}}^T\mathbf{A}{\mathbf{H}^T}\left( {\mathbf{G\tilde n + }{{\mathbf{\tilde n}}_\mathbf{a}}} \right)} \right|}^2}}} \\
\textit{s.t.} {\ \ \ \ }
& \max|\lambda_{{\mathbf{H}^T}\mathbf{G}}|<1, \notag\\
& G_i\geq0,{\ }i=1,\dots, N \notag
\end{align}
In scenarios where multiple EAs can receive signals from the station, the objective function can be written as the weighted summation over (\ref{snr_sa}), as shown in (\ref{sum_obj_fun}).
\begin{align}
\label{sum_obj_fun}
\mathop {\max}\limits_{\bf{G}} \text{SNR}_{\text{SA}}
&\Leftrightarrow \mathop {\max }\limits_{\bf{G}} \frac{{\sum\limits_i {{{\left| {{\bf{w}}_{\text{EA}}^T{\bf{A}}\left( {{\bf{w}}_{\text{SE}(i)}} {\tilde v}_i \right)} \right|}^2}}}}{{{{\left| {{\bf{w}}_{\text{EA}}^T{\bf{A}}{{\bf{H}}^T}\left( {{\bf{G\tilde n}} + {{{\bf{\tilde n}}}_{\bf{a}}}} \right)} \right|}^2}}}  \notag\\
&\Leftrightarrow \mathop {\max }\limits_{\bf{G}} \frac{{{{\left| {{\bf{w}}_{\text{EA}}^T{\bf{A}}\left(\sum\limits_i {{{\bf{w}}_{\text{SE}(i)}}} {\tilde v}_i \right) } \right|}^2}}}{{{{\left| {{\bf{w}}_{\text{EA}}^T{\bf{A}}{{\bf{H}}^T}\left( {{\bf{G\tilde n}} + {{{\bf{\tilde n}}}_{\bf{a}}}} \right)} \right|}^2}}} \\
\textit{s.t.} {\ \ \ \ }
& \max|\lambda_{{\mathbf{H}^T}\mathbf{G}}|<1, \notag\\
& G_i\geq0,{\ }i=1,\dots, N \notag
\end{align}
The deduction in (\ref{sum_obj_fun}) holds because the product of standard bases ${{{\bf{w}}_{\text{SE}(i)}}}{{{\bf{w}}_{\text{SE}(j)}^H}}=0 \text{, where } i\neq j$.

\subsection{Multiple-BSS Scenario}
\begin{figure}
\vspace{-0.3cm}
\centering
\includegraphics[width=0.48\textwidth]{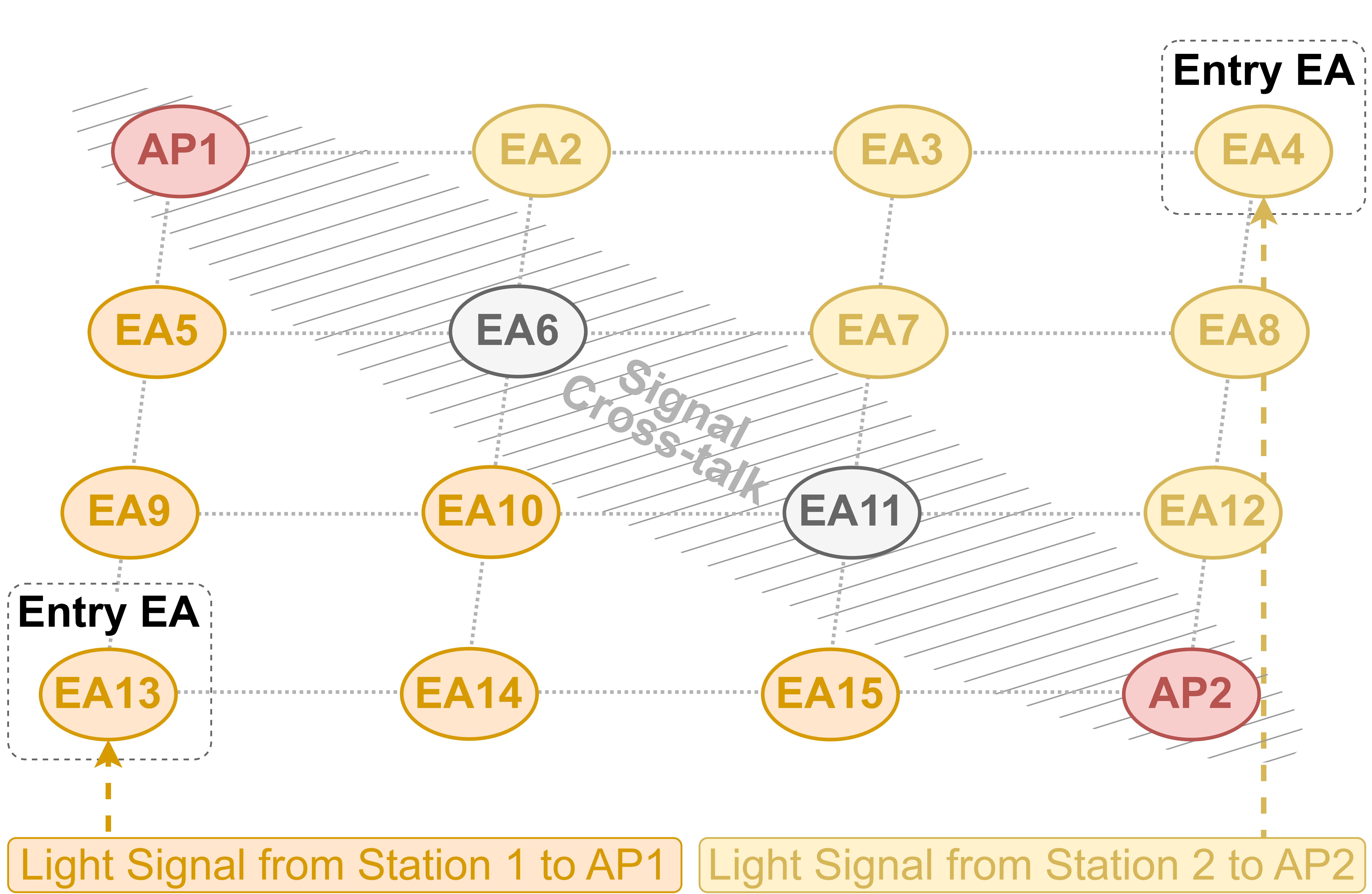}
\caption{A multiple-BSS scenario with two simultaneous communication links (EA13 $\rightarrow$ AP1 and EA4 $\rightarrow$ AP2). The EAs involved in signal propagation for each BSS are colored in orange and yellow, respectively.}
\label{spatial_reuse}
\vspace{-0.3cm}
\end{figure}

In multiple-BSS scenarios, each AP serves a distinct set of stations independently. Signal streams from different BSSs can propagate simultaneously through a shared OWE. However, as illustrated in Fig. \ref{spatial_reuse}, signal crosstalk occurs at the boundaries of adjacent BSSs (highlighted in grey), where overlapping transmissions from APs or stations create interference, degrading signal quality in both BSSs. Unlike the single-BSS scenario, the objective function in the multiple-BSS case shifts to maximize the signal-to-interference-plus-noise ratio (SINR) of signals traversing the OWE, mitigating \textbf{mutual interference} between BSSs. Furthermore, since BSSs share resources within the OWE, ensuring \textbf{fair resource allocation} is essential to maintain optimal performance across all BSSs.

\subsubsection{Mutual Interference Analysis}
The signal transmitted by a station in the \textit{j}-th BSS may be received by the AP in the \textit{i}-th BSS ($i\neq j$), causing interference from the \textit{j}-th BSS to the \textit{i}-th BSS. This mutual interference can be expressed as
\begin{equation}
\label{mutual_interference}
I_{m[j,i]}=\mathbf{\tilde y}_{[j]}^T w_{\text{EA}[i]}={\tilde v}_j{{\bf{w}}_{\text{SE}[j]}^T}\mathbf{A}^Tw_{\text{EA}[i]} \text{,} \quad i\neq j \text{ ,}
\end{equation}
where the subscripts [\textit{i}] and [\textit{j}] denote different BSSs and $\mathbf{\tilde y}_{[j]}$ is expanded as in (\ref{y_expansion}), with the noise component eliminated for pure interference analysis:
\begin{equation}
\label{y_expansion}
\mathbf{\tilde y}_{[j]} = \mathbf{A}\mathbf{\tilde x}_{[j]} \Rightarrow \mathbf{A} {{\mathbf{\tilde h}}_{\text{SE}[j]}}{{{\tilde s}_{\text{S}[j]}}} \Rightarrow \mathbf{A}{{\bf{w}}_{\text{SE}[j]}}{\tilde v}_j \text{ .}
\end{equation}
$\mathbf{\tilde y}_{[j]}$ is the vector of the received signal strength at EAs, corresponding to the signal (${{\tilde s}_{\text{S}[j]}}$) sent by the station in the \textit{j}-th BSS. 

\subsubsection{Fairness}
We assume all signal streams across multiple BSSs as equally important, irrespective of their signal strength or quality. Fairness, in this context, entails balanced resource allocation to prevent any stream from suffering substantial degradation relative to its performance in a single-BSS scenario.  Achieving this requires the joint optimization of SINRs across all BSSs.
\par
Therefore, a naive summation of SINRs across BSSs is insufficient. Due to inherent disparities in link quality -- such as stations closer to the AP achieving higher SINRs -- this approach may bias optimization toward certain BSSs, leading to suboptimal solutions where some streams dominate others unfairly.
To address this imbalance, we normalize each signal’s SINR by its corresponding optimal SNR in the single-BSS scenario, denoted as $\gamma_i^*=\max \text{SNR}_{\text{S}_i\text{A}}$, derived from optimizing the single-BSS objective function (\ref{snr_sa}).
\par
In cases where initial SNR disparities among BSS streams are extreme -- for instance, when one station is far from the AP while another is very close, resulting in SNR differences spanning orders of magnitude -- normalization alone may not suffice to correct the imbalance. To further mitigate this, we take the reciprocal of each SINR term in the summation. This ensures that low-SINR streams (where SINR approaches zero) are prioritized during optimization, as their reciprocal values grow rapidly, counteracting the dominance of high-SINR streams.
\par
Consequently, the objective function for optimizing EA gains in the multi-BSS scenario is formulated as
\begin{align}
\label{r_multi_bss_obj}
&\min_\mathbf{G} \sum_i \frac{\gamma_i^*}{\text{SINR}_{[i]}} \\
& = \min_\mathbf{G} \sum_i
\frac{\gamma_i^*\left[{{{\left| {\mathbf{w}_{\text{EA}[i]}^T\mathbf{A}{\mathbf{H}^T}\left( {\mathbf{G\tilde n + }{{\mathbf{\tilde n}}_\mathbf{a}}} \right)} \right|}^2}} + \sum_{j\neq i}\left|I_{m[j,i]}\right|^2\right]}{{{{|\tilde{v}_i|^2\left|\mathbf{w}_{\text{EA}[i]}^T\mathbf{A}{\mathbf{w}_{\text{SE}[i]}} \right|}^2}}} \notag \\
&\textit{s.t.} {\ \ \ \ }
\max|\lambda_{{\mathbf{H}^T}\mathbf{G}}|<1, \notag\\
& {\ \ \ \ \ \ \ \ } G_i\geq0,{\ }i=1,\dots, N \text{ .} \notag
\end{align}
\section{Simulations}
To evaluate OWE's capability in extending optical signal coverage and demonstrate the performance enhancement achieved through OWE channel optimization (Section III), we performed comprehensive simulations.
\subsection{Signal Coverage Extension Capability}
\begin{figure}
\centering
\includegraphics[width=0.48\textwidth]{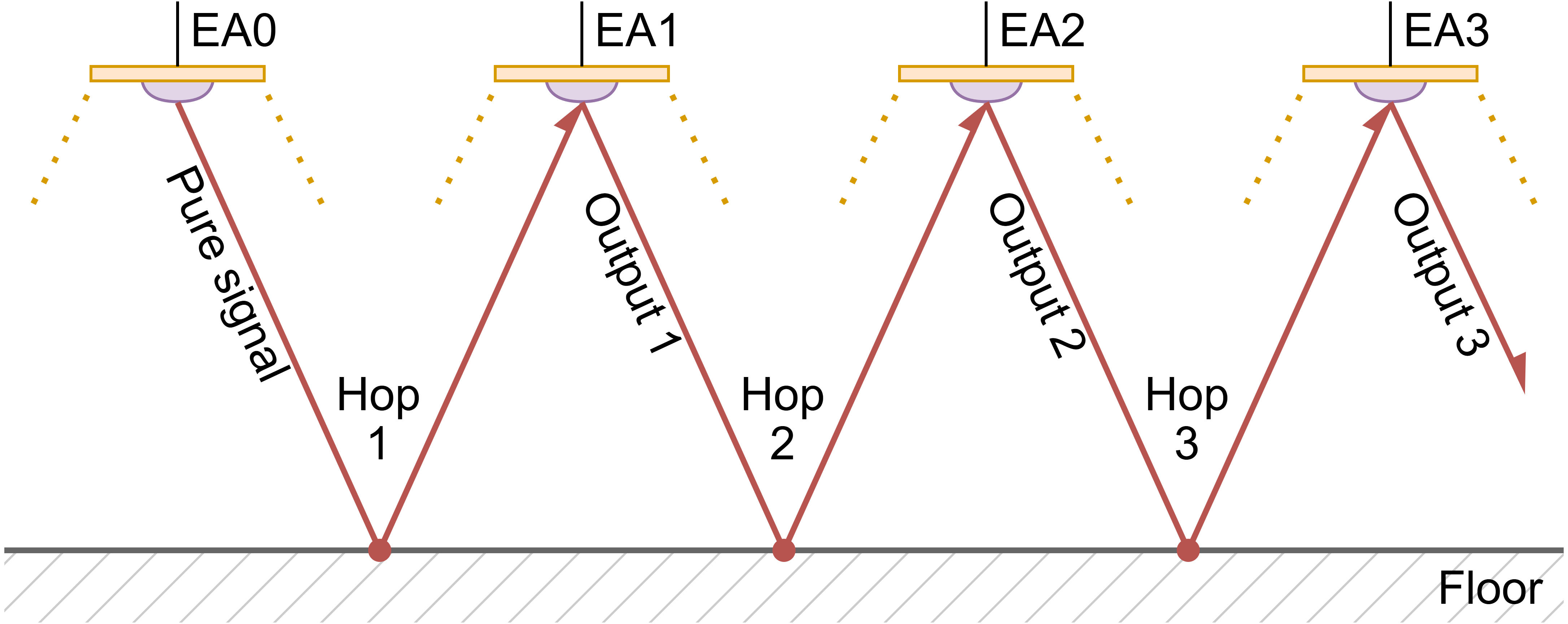}
\caption{Simulation setup for evaluating signal coverage extension capability.}
\label{coverage_simulation}
\vspace{-0.5cm}
\end{figure}

In OWE systems, signal quality gradually degrades during propagation due to attenuation in the diffuse reflection channel and noise introduced at each EA. We evaluated this effect by simulating signal degradation along a straight-line propagation path (Fig. \ref{coverage_simulation}). The simulation assumed that EA0 emits a noise-free optical signal of $8.1 \text{ W}$. The degradation rate depends on the EA gain settings (particularly the PA gain), where proper amplification helps mitigate signal loss. Additional simulation parameters are listed in Table \ref{simulation_settings}.
\begin{table*}
\centering
\begin{tabularx}{1\textwidth}{>{\raggedright}p{6.8cm} l | >{\raggedright}p{6.8cm} l}
\toprule
\textbf{Parameters}&\textbf{Values}&\textbf{Parameters}&\textbf{Values}\\ \midrule 
Room ceiling height&$2.5\text{\ m}$                             &PD shunt resistance&$100 \text{\ M}\Omega$\\ \midrule 
Horizontal install distance between EAs&$1.25\text{\ m}$        &Constant background optical power&$1\ \mu \text{W}$\\ \midrule
Floor reflectivity, assuming a Lambertian reflector&$0.4$       &\textbf{PD RMS noise}$^{(\text{iii})}$ (calculated at $P_r=52 \ \mu\text{W}$)&$13 \text{\ nA}$\\ \midrule
System bandwidth&$20 \text{\ MHz}$                              &TIA feedback resistor ($R_{f(\text{TIA})}$)&$10\text{\ k}\Omega$\\ \midrule
LED radiant efficiency&$0.45$                                   &TIA input voltage noise density&$1{\text{\ nV}/\sqrt{\text{Hz}}}$\\ \midrule
LED electrical power&$18 \text{\ W}$                            &TIA input current noise density&$2.5{\text{\ pA}/\sqrt{\text{Hz}}}$\\ \midrule
LED luminous efficacy&$80 \text{\ lm/W}$                        &\textbf{TIA RMS noise}$^{(\text{iii})}$ (referred to input; calculated at $P_r=52 \ \mu\text{W}$)&$18.1 \text{\ nA}$\\ \midrule
LED forward voltage&$3 \text{\ V}$                              &PA input resistance ($R_{i(\text{PA})}$)&$50\ \Omega$\\ \midrule
Brightness of a $5\times 5 \times 2.5 \text{\ m}^3$ room illuminated by a $3\times3$ EA grid network (meet lighting regulations)&$518.4 \text{\ lux}$&PA feedback resistance ($R_{f(\text{PA})}$; adjustable for EA programmable gain)&$2\text{\ K}\Omega$\\ \midrule
Half power angle of the optical transmitter $^{(\text{i})}$ ($\Psi_{1/2}$)&$19.65\degree$&PA input voltage noise density &$3{\text{\ nV}/\sqrt{\text{Hz}}}$\\ \midrule
PD receiving area ($A_r$)&$10^{-4}\text{\ m}^2$                 &PA input current noise density &$10{\text{\ pA}/\sqrt{\text{Hz}}}$\\ \midrule 
Acceptance angle of the optical receiver $^{(\text{ii})}$ ($\Psi_C$)&$34.21\degree$&\textbf{PA RMS noise} (referred to TIA input)&$1.47\text{\ nA}$\\ \midrule
Refractive index of concentrator before the PD ($\tau$)&$1.5$   &Bias tee current limiting resistor ($R_{\text{BT}}$)&$5\ \Omega$\\ \midrule 
PD responsivity ($r$)&$0.5 \text{\ A/W}$                        &Bias tee inductor&$10 \ \mu\text{H}$\\ \midrule 
PD dark current&$2 \text{\ nA}$                                 &DC-DC converter RMS noise (going through the LED)&$0.42 \text{\ mA}$\\
\bottomrule
\end{tabularx}
\caption{Simulation parameter settings. Note: (i)-(ii) The rationale and calculations for these configurations are detailed in Appendix B. (iii) Parameter values are interdependent and may vary based on other parameter settings.}
\label{simulation_settings}
\vspace{-0.3cm}
\end{table*}
\par
\begin{figure}
\centering
\includegraphics[width=0.48\textwidth]{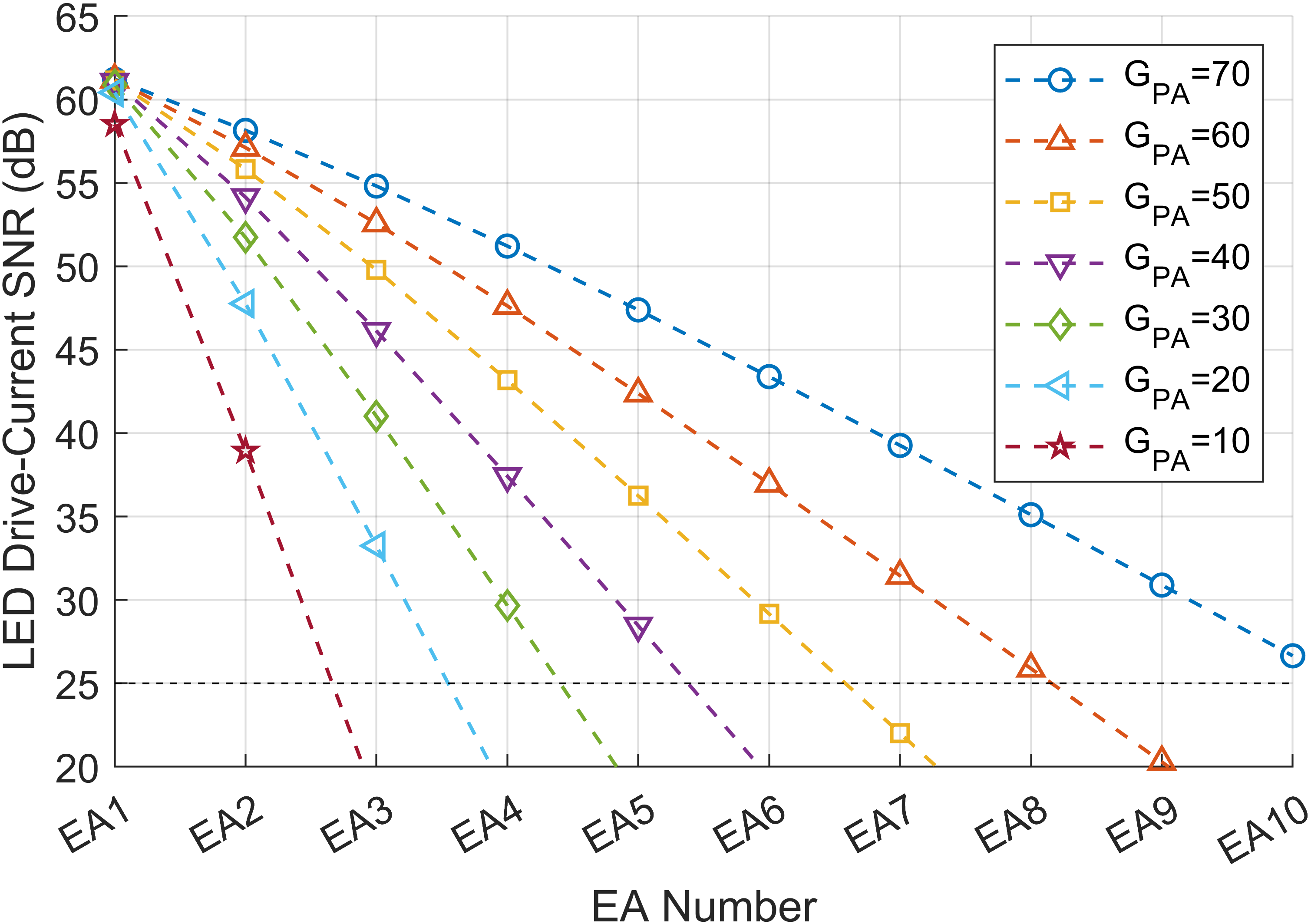}
\caption{Simulation results of the signal coverage extension capability with different EAs gain settings.}
\label{coverage_simulation_res}
\vspace{-0.5cm}
\end{figure}

Figure \ref{coverage_simulation_res} shows the LED drive-current SNRs across EA1 to EA10. With the PA gain of all EAs set to 70, the SNR remains above 25 dB over 10 hops, covering a horizontal propagation distance of 12.5 m.
This result confirms that OWE can provide signal coverage for large indoor spaces such as airports or supermarkets. To further optimize deployment, the transmit power of individual EAs can be increased to extend the propagation distance per hop, thereby reducing the total number of EAs required.
Even at a reduced PA gain of 10, the signal can propagate two hops, sufficient to traverse a 3$\times$3 EA network, ensuring coverage for a 5$\times$5 $\text{m}^2$ room (the grid network is positioned 1.25 m from the walls, and EAs in which are uniformly spaced at 1.25 m intervals).

\subsection{Single-BSS Scenario}
For the single-BSS scenario, we optimize the objective function defined in (\ref{snr_sa}) to maximize the SNR of signals propagating through the OWE. Since (\ref{snr_sa}) is non-convex, we employ the simulated annealing method to avoid convergence to local optima. The pseudo-code for the optimization algorithm is provided in Algorithm \ref{simulated_annealing}.
\begin{algorithm}
\caption{Simulated Annealing}
\label{simulated_annealing}
\begin{algorithmic}[1]
\State \textbf{Input:} Initial EA gain settings $ \mathbf{G}_0 $, initial temperature $ T_0 $, cooling rate $ \alpha $, minimum temperature $ T_{\text{min}} $, maximum iterations $ \text{max\_iter} $
\State \textbf{Output:} Best solution $ \mathbf{G}_{\text{best}} $
\State $ \mathbf{G} \gets \mathbf{G}_0 $, $ \mathbf{G}_{\text{best}} \gets \mathbf{G} $, $ T \gets T_0 $, $ \text{iter} \gets 0 $
\State $f(\mathbf{G})\gets -\text{SNR}_{\text{SA}}(\mathbf{\mathbf{G}}) \text{\ (defined in (\ref{snr_sa}))}$ 

\While{$ T > T_{\text{min}} $ \textbf{and} $ \text{iter} < \text{max\_iter} $}
    \State Generate a neighboring solution $ \mathbf{G}' $ randomly from $ \mathbf{G} $
    \State Compute $ \Delta f =f(\mathbf{G'}) -f(\mathbf{G}) $
    
    \If{$ \Delta f < 0 $ \textbf{or} $ \text{random}(0,1) < e^{-\Delta f / T} $}
        \State $ \mathbf{G} \gets \mathbf{G}' $ \Comment{Accept the new gain settings}
        \If{$ f(\mathbf{G}) < f(\mathbf{G}_{\text{best}}) $}
            \State $ \mathbf{G}_{\text{best}} \gets \mathbf{G} $
        \EndIf
    \EndIf
    
    \State $ T \gets \alpha \cdot T $ \Comment{Cooling schedule}
    \State $ \text{iter} \gets \text{iter} + 1 $
\EndWhile

\State \textbf{return} $ \mathbf{G}_{\text{best}} $
\end{algorithmic}
\end{algorithm}
\par
We investigated a 3$\times$3 EA grid network deployed in a 5$\times$5 $\text{m}^2$ room, with other parameters configured as specified in Table \ref{simulation_settings}. To thoroughly evaluate OWE performance and account for the random placement of stations, we selected EAs 1, 2, 3, 5, 6, and 9 as representative signal entry EAs. The simulation results for the other EAs can be easily inferred from the symmetric structure of the network. We assumed that the optical signal emitted by the station is noiseless and reaches the entry EA's PD with a power of $52\ \mu \text{W}$. Notably, EA9 is directly connected to the AP as depicted in Fig. \ref{ea_ap}.
\begin{figure}[htbp]
\vspace{-0.2cm}
\centering
\includegraphics[width=0.5\textwidth]{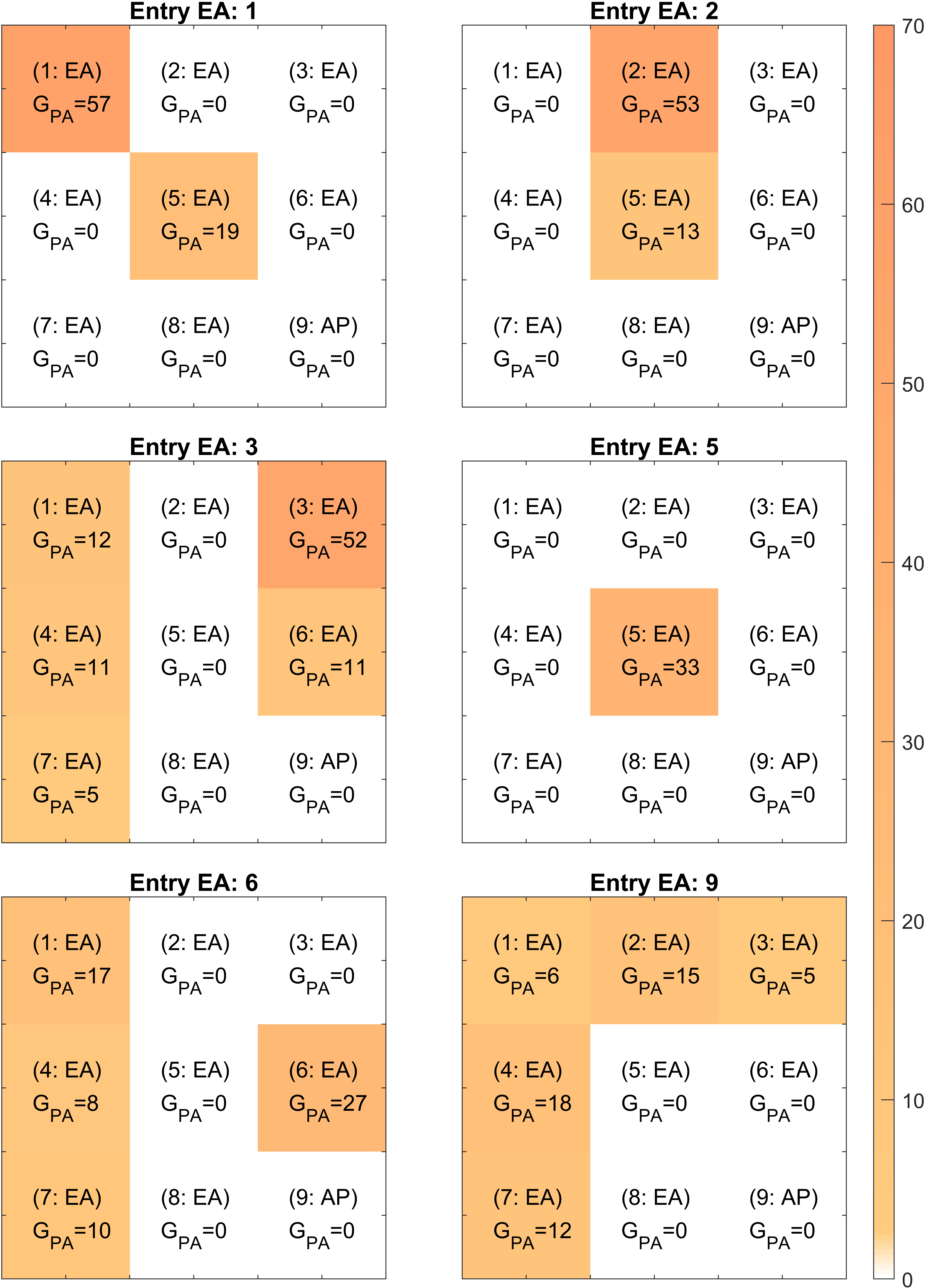}
\caption{Optimization results for OWE single-BSS application scenarios. 6 EAs are selected individually as the entry EA. EA 9 is integrated with the AP.}
\label{single_bss_sim_res}
\end{figure}
\par
Figure \ref{single_bss_sim_res} presents the optimization results, with each subfigure labeled by its corresponding entry EA. As shown, EA9 constantly exhibits a gain value of 0, whereas the entry EAs are always assigned the highest gain value among other EAs. This result is justified because amplification not only enhances the signal but also magnifies embedded noise while introducing additional noise components. By applying higher gain at the initial OWE propagation stage -- when noise levels are minimal -- rather than at later stages where noise accumulates, the system improves the SNR at the AP.
\par
In particular, when EA9 serves as the entry EA, directly receiving the signal without EA amplification maximizes the SNR at the AP. This gain setting avoids the gain-dependent noise introduced by the TIA and PA. Furthermore, EA5, EA6, and EA8, which surround EA9, were optimized to a gain of 0, effectively isolating EA9 from other EAs to minimize noise.
\par
The improvements in SNR after deploying our optimization are shown in Table \ref{snr_improvement}. Notably, the improvement is relative to the equal gain setting, where all EAs share the same maximum gain value that fulfills the stability constraints. The EA9 case achieved a significant SNR improvement due to the near-zero noise introduced by deploying optimized gain settings and the noiseless assumption of the station signal. 
\par
Overall, the simulation results demonstrate that in single-BSS scenarios, optimizing the EA gain settings using our proposed objective function significantly improves signal quality in OWE. Additionally, the system exhibits adaptivity to dynamic environments with varying station positions.

\begin{table}
\centering
\begin{tabularx}{0.49\textwidth}{p{1.2cm}l|p{1.2cm}l}
\toprule
\begin{tabular}[c]{@{}l@{}} \textbf{Entry} \\ \textbf{EA} \end{tabular} & \begin{tabular}[c]{@{}l@{}} \textbf{SNR} \\ \textbf{[Improvement]} \quad \quad \quad \end{tabular} & \begin{tabular}[c]{@{}l@{}} \textbf{Entry} \\ \textbf{EA} \end{tabular} & \begin{tabular}[c]{@{}l@{}} \textbf{SNR} \\ \textbf{[Improvement]} \end{tabular} \\ \midrule
1           & $ \text{50.19\ dB}\ [\text{14.34\ dB}]$    & 2           & $ \text{50.81\ dB}\ [\text{12.85\ dB}]$   \\
3           & $ \text{51.47\ dB}\ [\text{15.08\ dB}]$    & 5           & $ \text{57.11\ dB}\ [\text{16.69\ dB}]$    \\
6           & $ \text{57.99\ dB}\ [\text{18.56\ dB}]$    & 9           & $ \text{84.64\ dB}\ [\text{42.17\ dB}]$    \\ \bottomrule
\end{tabularx}
\caption{SNR values and corresponding improvements achieved by applying the optimized gain settings in Fig. \ref{single_bss_sim_res}. The improvements are computed relative to the equal gain settings with the same entry EA, respectively.}
\label{snr_improvement}
\vspace{-0.4cm}
\end{table}
\subsection{Multiple-BSS Scenario}
For the multiple-BSS scenario, we optimize the EA gain values using the objective function in (\ref{r_multi_bss_obj}) via simulated annealing, following similar pseudocode as Algorithm \ref{simulated_annealing}. The only modification is in line 4, where we replace the objective with $f(\mathbf{G})\leftarrow\sum_i \frac{\gamma_i^*}{\text{SINR}_{[i]}} \text{\ (defined in (\ref{r_multi_bss_obj}))}$.
\par
Our simulations were conducted in a 4$\times$4 grid OWE with 1.25 m spacing between EAs, deployed in a 6.25$\times$6.25 $\text{m}^2$ room. Other parameters followed the settings specified in Table \ref{simulation_settings}. We initially assumed that the same optical power of $52\ \mu \text{W}$ reaches the PDs of entry EAs in different BSSs. However, variations in incident signal power at these entry EAs influence mutual interference, SINR, and fairness in resource allocation among BSSs -- a topic we will explore in detail later.
We evaluated four multiple-BSS scenarios where the OWE system simultaneously propagated signals from two entry EAs to two distinct APs (AP1 and AP2).
\begin{figure*}
\vspace{-0.3cm}
\centering
\includegraphics[width=1\textwidth]{Figures/Multi_BSS_SINR_5.png}
\caption{OWE Multiple-BSSs application optimization results. Two data streams propagate from two Entry EAs through the OWE to two APs marked as AP1 and AP2 simultaneously.}
\label{multi_bss_sim_f}
\end{figure*}

\begin{table*}[htbp]
\centering
\begin{tabularx}{1\textwidth}{p{5.5cm}|XXXX}
\toprule
\textbf{\begin{tabular}[c]{@{}l@{}}Signal Paths \\ (Entry EA $\rightarrow$ AP $^{(\text{i})}$)\end{tabular}}                                                                   & \textbf{\begin{tabular}[c]{@{}l@{}}EA13 $\rightarrow$ AP1 \\ EA4 $\rightarrow$ AP16\end{tabular}} & \textbf{\begin{tabular}[c]{@{}l@{}}EA1 $\rightarrow$ AP1 \\ EA4 $\rightarrow$ AP2\end{tabular}} & \textbf{\begin{tabular}[c]{@{}l@{}}EA10 $\rightarrow$ AP1 \\ EA1 $\rightarrow$ AP2\end{tabular}} & \textbf{\begin{tabular}[c]{@{}l@{}}EA1 $\rightarrow$ AP1 \\ EA3 $\rightarrow$ AP2\end{tabular}} \\ \midrule
\textbf{AP1 Received Signal SINR [Degradation $^{(\text{ii})}$]} & $46.89 \text{\ dB}\  [-0.01 \text{\ dB}]$     & $46.90 \text{\ dB} \ [0.00 \text{\ dB}]$  & $57.46 \text{\ dB} \ [0.00 \text{\ dB}]$ & $41.22 \text{\ dB} \ [-4.82 \text{\ dB}]$ \\ \midrule
\textbf{AP2 Received Signal SINR [Degradation]} & $46.90 \text{\ dB} \ [0.00 \text{\ dB}]$     & $46.90 \text{\ dB} \ [0.00 \text{\ dB}]$ & $41.15 \text{\ dB} \ [-4.36 \text{\ dB}]$ & $43.73 \text{\ dB} \ [-2.85 \text{\ dB}]$ \\ \midrule
\textbf{AP1 Mutual Interference Power Ratio $^{(\text{iii})}$}  & 0     & 0 & 0 & $1.6\times 10^{-5}$ \\ \midrule
\textbf{AP2 Mutual Interference Power Ratio}  & 0     & 0 & 0 & 0 \\ \bottomrule
\end{tabularx}
\caption{The simulation results of the signal quality and mutual interference in the multiple-BSS scenarios.
Note: (i) Across simulation scenarios, the AP positions vary and correspond to Figure \ref{multi_bss_sim_f} from left to right. (ii) $\text{Degradation}= 10\log_{10}\left(\frac{\text{Multiple-BSS SINR}}{\text{Single-BSS SNR}}\right)$. (iii) $\text{Mutual Interference Power Ratio}= \frac{\text{Interference Power}}{\text{Received Signal Power}}$.}
\label{multi_bss_sim_t}
\vspace{-0.5cm}
\end{table*}
\par
The simulation results, presented in Fig. \ref{multi_bss_sim_f} and Table \ref{multi_bss_sim_t}, demonstrate that: 1) The OWE system successfully maintains concurrent operation of multiple BSSs while preserving each BSS's performance, effectively addressing fairness concerns in resource allocation. 2) Different BSSs remain effectively isolated, with mutual interference being well suppressed.
\par
A notable example is the third simulation case, which analyzed two signal streams: EA1 $\rightarrow$ AP2 and EA10 $\rightarrow$ AP1. Due to their different propagation distances, the optimal SNR values for these streams varied significantly. The simulation results show that our objective function can jointly optimize the unbalanced elements (in the summation, corresponding to BSSs) towards their respective optima, effectively addressing fairness concerns. In the fourth case: EA1 $\rightarrow$ AP1 and EA3 $\rightarrow$ AP2, we tested an extreme scenario with closely placed APs. Although mutual interference was unavoidable in this configuration, our approach successfully optimized light propagation routing and effectively suppressed the impact of interference.
\begin{figure}
\centering
\includegraphics[width=0.5\textwidth]{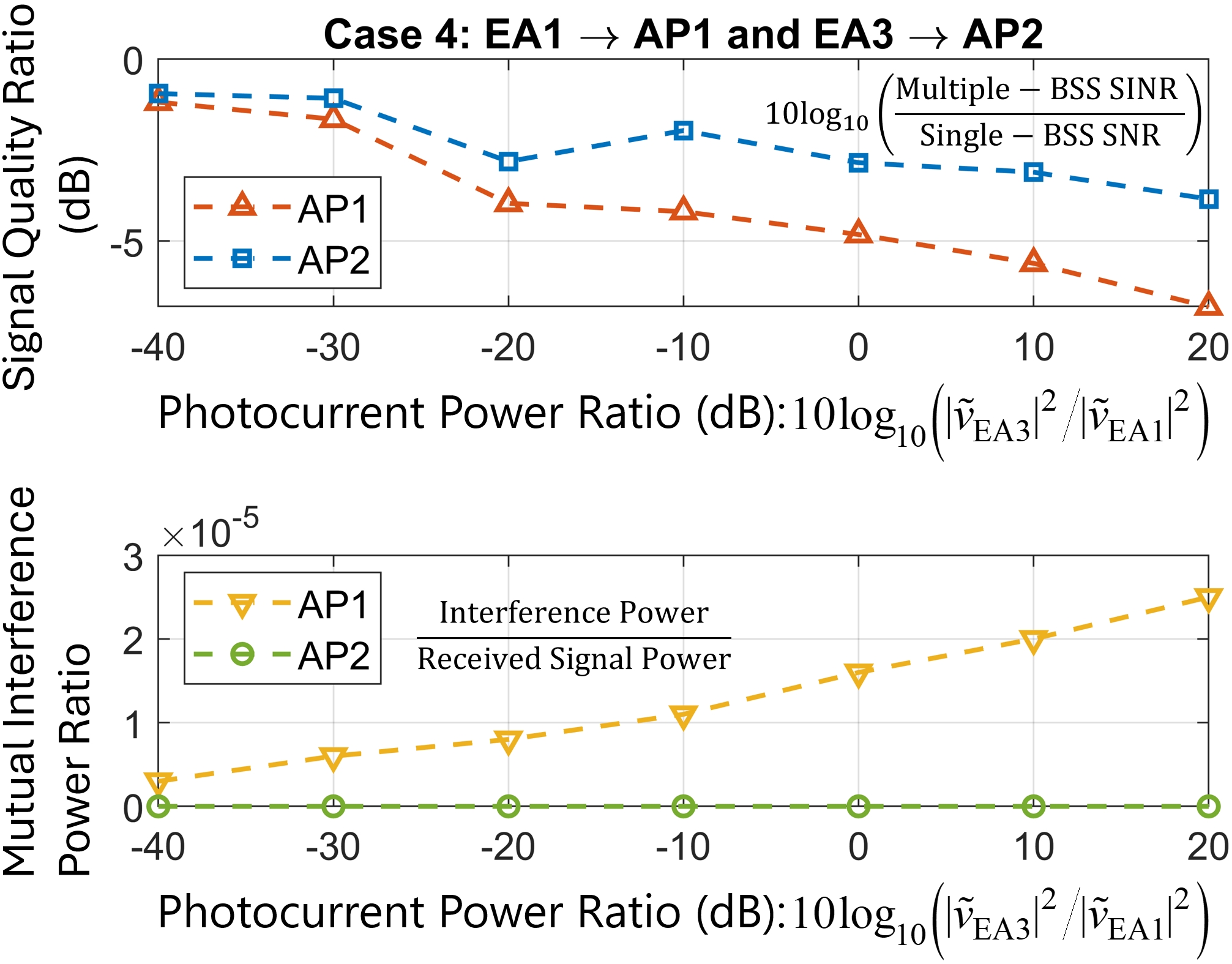}
\caption{Optimization results for the fourth case of OWE multiple-BSS application scenarios in Fig. \ref{multi_bss_sim_f} and Table \ref{multi_bss_sim_t}, obtained by varying the pre-gain SNR at the entry EA.}
\label{psi_influence}
\vspace{-0.5cm}
\end{figure}
\par
Considering that the incident optical signal power at entry EAs is typically variable and unequal, we investigated how this imbalance affects EA gain optimization in multi-BSS scenarios. Using the fourth case in Fig. \ref{multi_bss_sim_f} as a representative example, we fixed the photocurrent amplitude of EA1 at $\tilde{v}_{\text{EA1}}=26 \ \mu \text{A}$ (corresponding to $52\ \mu \text{W}$ of optical power incident on the PD) and varied $\tilde{v}_{\text{EA3}}$ to cover a broad range of photocurrent ratios ($|{\tilde{v}_{\text{EA3}}}|/|{\tilde{v}_{\text{EA1}}|}$ from 0.01 to 10).
\par
The simulation results in Fig. \ref{psi_influence} demonstrate that when the photocurrent at EA3 is just $1\%$ of that at EA1, signal degradation and mutual interference are minimal. As $|\tilde{v}_{\text{EA3}}|^2$ increases by 60 dB, the SINR degradation reaches 5.6 dB for AP1 and 2.9 dB for AP2. This arises from the trade-off in eliminating the mutual interference. Despite this, the degradation is relatively small compared to the wide range of photocurrent power, confirming that our approach effectively handles signal power variations while maintaining fairness across all signal streams in multi-BSS scenarios.
\par
In summary, the simulation results validate that our optimization approach effectively resolves both interference and fairness issues in multiple-BSS scenarios, consistent with our theoretical analysis. The system exhibits robust performance under variations in station/AP positioning and differences in incident optical signal power across BSSs.
\section{Real Experiments}
In addition to simulation studies, we developed a 4-EA (2$\times$2) prototype OWE and conducted real-world experiments to achieve two objectives:
1) Feasibility Validation:  Verify the practical feasibility by implementing OWE over a Wi-Fi-based TCP/IP network, measuring SNR improvement and packet loss reduction to confirm compatibility with existing technologies.
2) Additional Configuration Exploration: Investigate a multiple-EA reception scenario, where multiple EAs capture transmitted signals from the station (as defined in (\ref{sum_obj_fun})), in contrast to the single-EA setup described in (\ref{snr_sa}).
\subsection{Hardware Design and Experimental Setups}
\begin{figure}
\centering
\includegraphics[width=0.5\textwidth]{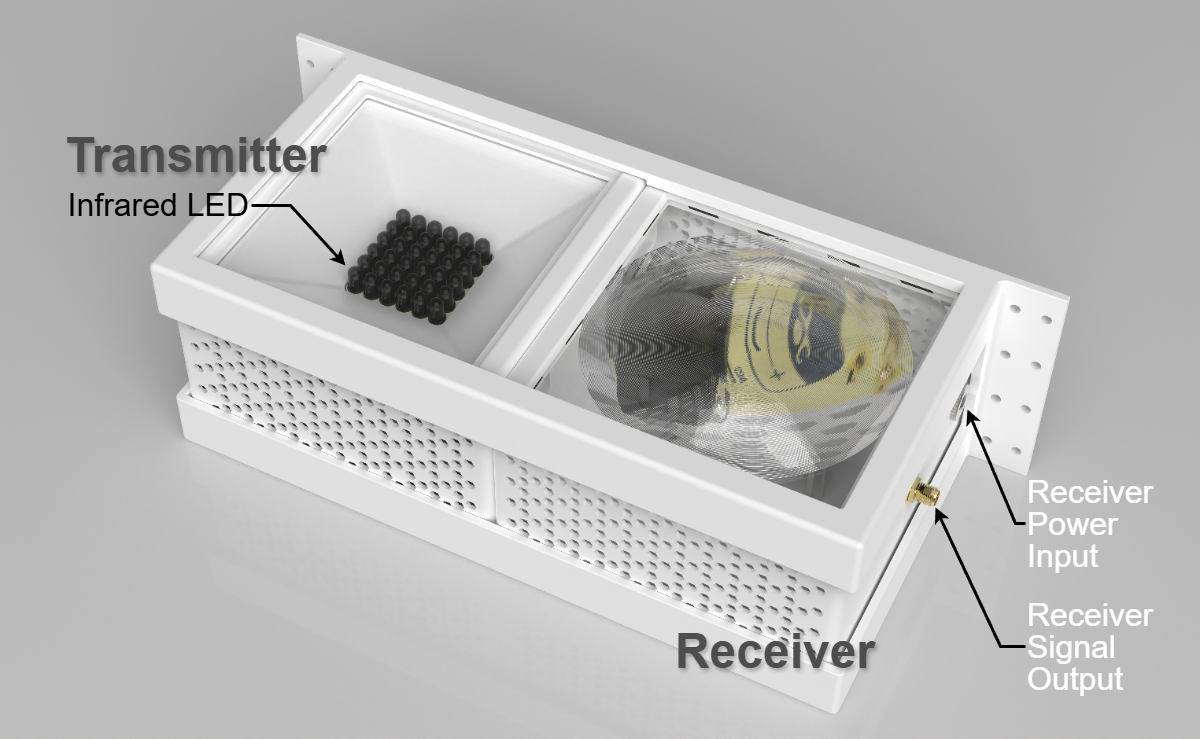}
\caption{The rendered image of the optical transceiver front ends integrated in EAs.}
\label{Transceiver_Rendered}
\vspace{-0.3cm}
\end{figure}

We designed and built EA hardware using commercial off-the-shelf components as depicted in Fig. \ref{Transceiver_Rendered}. The EA can provide a -3 dB bandwidth exceeding 40 MHz, with the TIA stage configured for a fixed gain of $10\ \text{k}\Omega$ and the PA gain adjustable from 0 (disabled) to 100. When the PA gain is set to 70 with no optical input signal, the measured voltage noise at the PA output is $9.5\ \text{mV}_{\text{RMS}}$. In the prototype system, the EA transmitter employs an infrared LED module with a total electrical power consumption of approximately $4 \text{ W}$. Since the system is designed solely for research purposes and uses invisible infrared light for communication, it does not provide illumination. Notably, the choice of wavelength does not impact the feasibility of OWE, as the system operates via intensity-modulated signals rather than direct modulation on the optical frequencies.
\footnote{For practical implementations requiring illumination, RGB LEDs could serve as dual-purpose light sources, enabling both communication and lighting. RGB LEDs produce white light through color mixing and inherently offer broader bandwidth compared to conventional phosphor-coated white LEDs \cite{bian201810}.}
For our experiments, to ensure accurate measurement of OWE-introduced noise, all ambient artificial lighting was turned off.
\par
\begin{figure}
\centering
\includegraphics[width=0.5\textwidth]{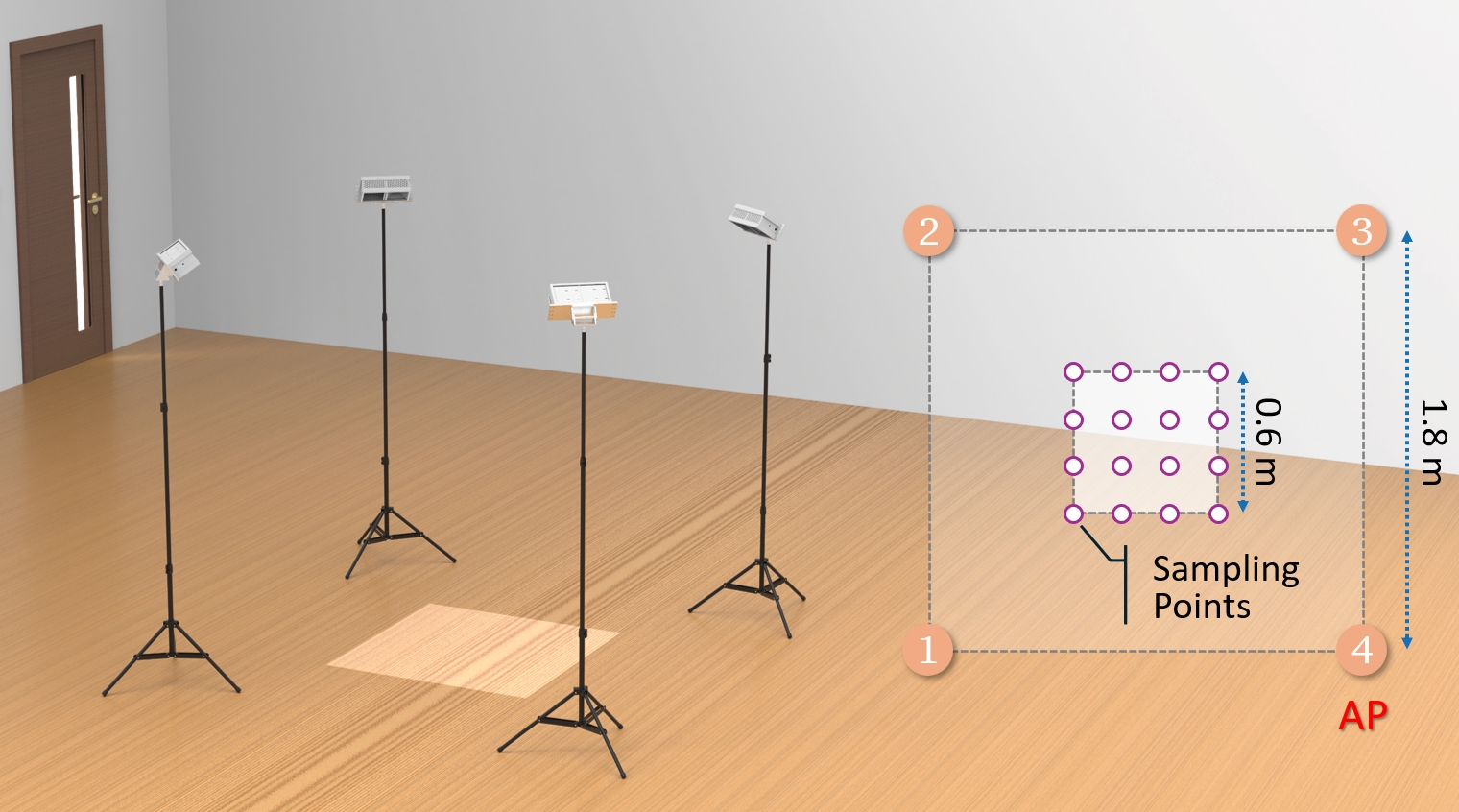}
\caption{Experimental scene and settings.}
\label{experimental_settings}
\vspace{-0.5cm}
\end{figure}

In the prototype OWE, 4 EAs were positioned at the corners of a 1.8$\times$1.8 $\text{m}^2$ square experimental area. As shown in Fig. \ref{experimental_settings}, the EA serving as the AP was placed at the bottom-right corner (top view). Each EA was mounted at a height of 2 m above the floor, with its optical front ends slightly tilted rather than perpendicular to the ground. 
For uplink signal propagation testing, a small optical transmitter with an electrical power of approximately 0.89 W acted as the station. The transmitter was placed on the floor within a 0.6$\times$0.6 $\text{m}^2$ square centered in the experimental area, radiating the optical signal upward. To evaluate performance, we selected 16 evenly distributed sampling points (marked by purple circles in Fig. \ref{experimental_settings}) and assessed two key metrics: SNR and packet loss rate.

\begin{figure*}[hbp]
\vspace{-0.5cm}
\centering
\subfloat[SNR measurements from 16 sampling points for three different EA gain settings.
Note: The grids in this figure represent different station positions used for SNR measurements (see Fig. \ref{experimental_settings}).]
{
  \label{snr_surf}
  \includegraphics[width=0.6792\textwidth]{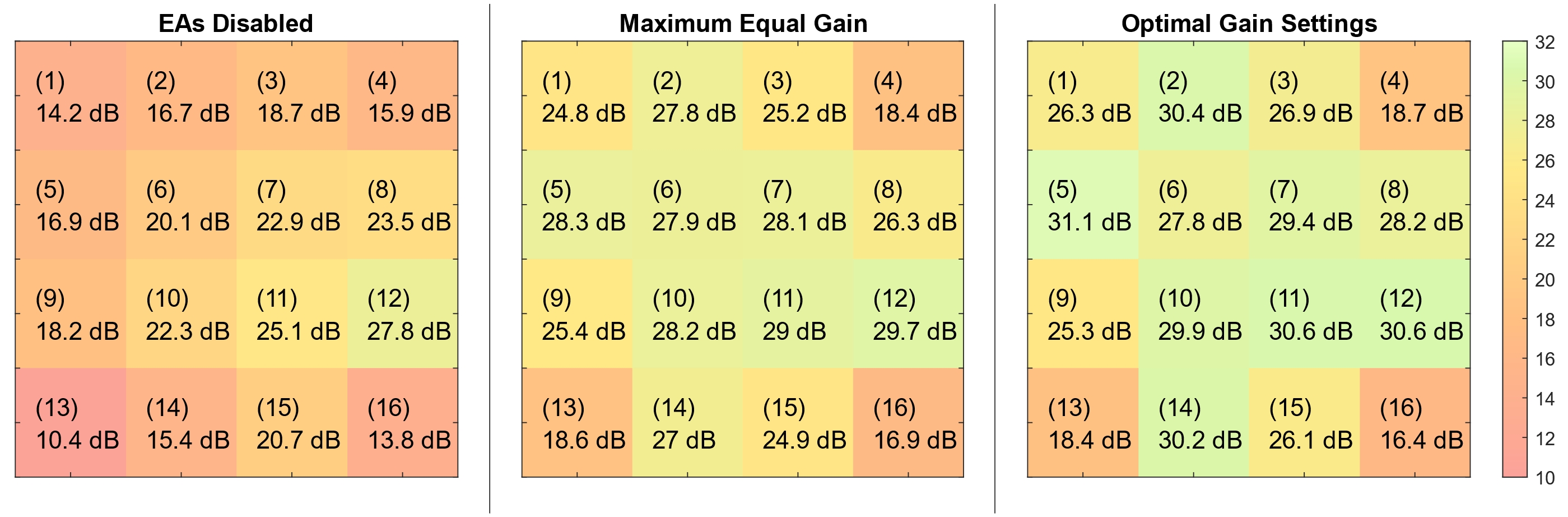}
}
\subfloat[SNR surf plot.]
{
  \label{snr_stack}
  \includegraphics[width=0.2707\textwidth]{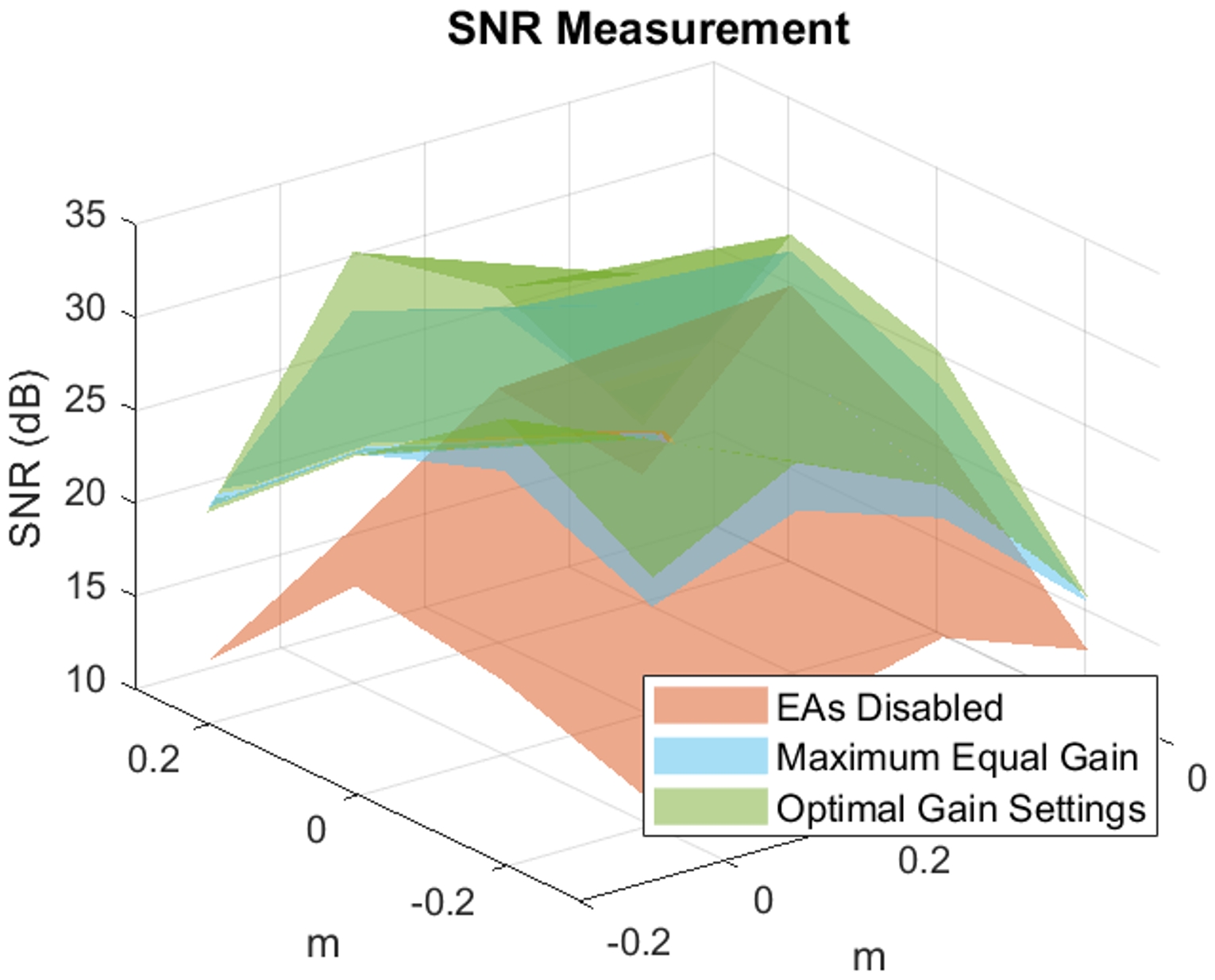}
}
\caption{SNR measurement results.}
\label{snr_experiments}
\vspace{-0.5cm}
\end{figure*}

\subsection{SNR}
In the SNR test experiments, the station continuously sent a 1 MHz sinusoidal waveform. We measured the SNR of the signal received by the AP under different EA gain settings.
\par
First, we collected SNR values with all EAs disabled, in which case the signal can only propagate through LOS paths. In a large-scale network, if the distance between the station and the AP is too long without any available LOS paths, the connection will be lost.
\par
Second, we explored the equal-gain setting, where we maximized a common gain across all EAs while satisfying non-saturation constraints to ensure overall OWE stability.
\par
Third, we applied the optimal gain settings derived by optimizing the objective function (\ref{sum_obj_fun}) using the same simulated annealing method employed in the simulations.

\begin{table}
\centering
\begin{tabularx}{0.48\textwidth}{p{1.8cm} p{1.7cm} X X}
\toprule
\textbf{Gain Settings} &
  \textbf{Disable EAs} &
  \textbf{\begin{tabular}[c]{@{}l@{}}Maximum\\ Equal Gain\end{tabular}} &
  \textbf{\begin{tabular}[c]{@{}l@{}}Optimal\\ Gain Settings\end{tabular}} \\ \midrule
\textbf{Average SNR} &
  21.20   dB &
  26.61   dB &
  28.25   dB \\ \bottomrule
\end{tabularx}
\caption{Average SNR of three EA gain settings.}
\label{average_snr}
\vspace{-0.5cm}
\end{table}
\par
As shown in Fig. \ref{snr_experiments} and Table \ref{average_snr}, the SNR of the received signal at the AP can be improved by applying proper EA gain settings, especially at positions with weak LOS paths such as sampling points numbered 1, 2, and 5.
Notably, although sampling point 16 is the closest to the AP (located at the bottom-right corner), the LOS signal is substantially attenuated due to the tilted installation of EAs, leading to degraded signal quality. Furthermore, because sampling point 16 is distant from -- and has only a weak connection to -- the other three EAs in the OWE, it derives minimal benefit from gain optimization.
\par
In summary, the SNR performance with the optimal EA gain settings surpasses that with maximum equal gain settings, which, in turn, outperforms the scenario with EAs disabled. A significant improvement of over 7 dB in average SNR is achieved by deploying the optimal EA gain settings compared to having EAs disabled.

\subsection{Packet Loss Rate}
In packet loss rate experiments, we used USRP \cite{ettus2015universal} to generate Wi-Fi packets with three modulations: 16-QAM 1/2 coding rate, 16-QAM 3/4 coding rate, and 64-QAM 1/2 coding rate. The USRP transmitted Wi-Fi signals at 2.4 GHz with a modulation bandwidth of 5 MHz. For each experimental trial, we generated and analyzed 4095 packets.
\par
Following the method introduced in \cite{cui2024wi}, we down-convert the 2.4 GHz RF signal to an intermediate frequency signal centered at 15 MHz, then convert it to an intensity-modulated optical signal at the transmitter. At the receiver, this process was reversed -- converting the optical signal back to an RF signal for capture by the USRP. In this configuration, OWE is resilient to ambient interference, as its modulated signal operates in the MHz range, far above the kHz-range flickering typical of ambient light. This frequency separation allows for straightforward noise filtering.

\begin{figure*}
\vspace{-0.5cm}
\centering
\subfloat[Packet loss rate over different sampling points, EA gain settings, and modulations.]
{
  \label{plr_surf}
  \includegraphics[width=0.715\textwidth]{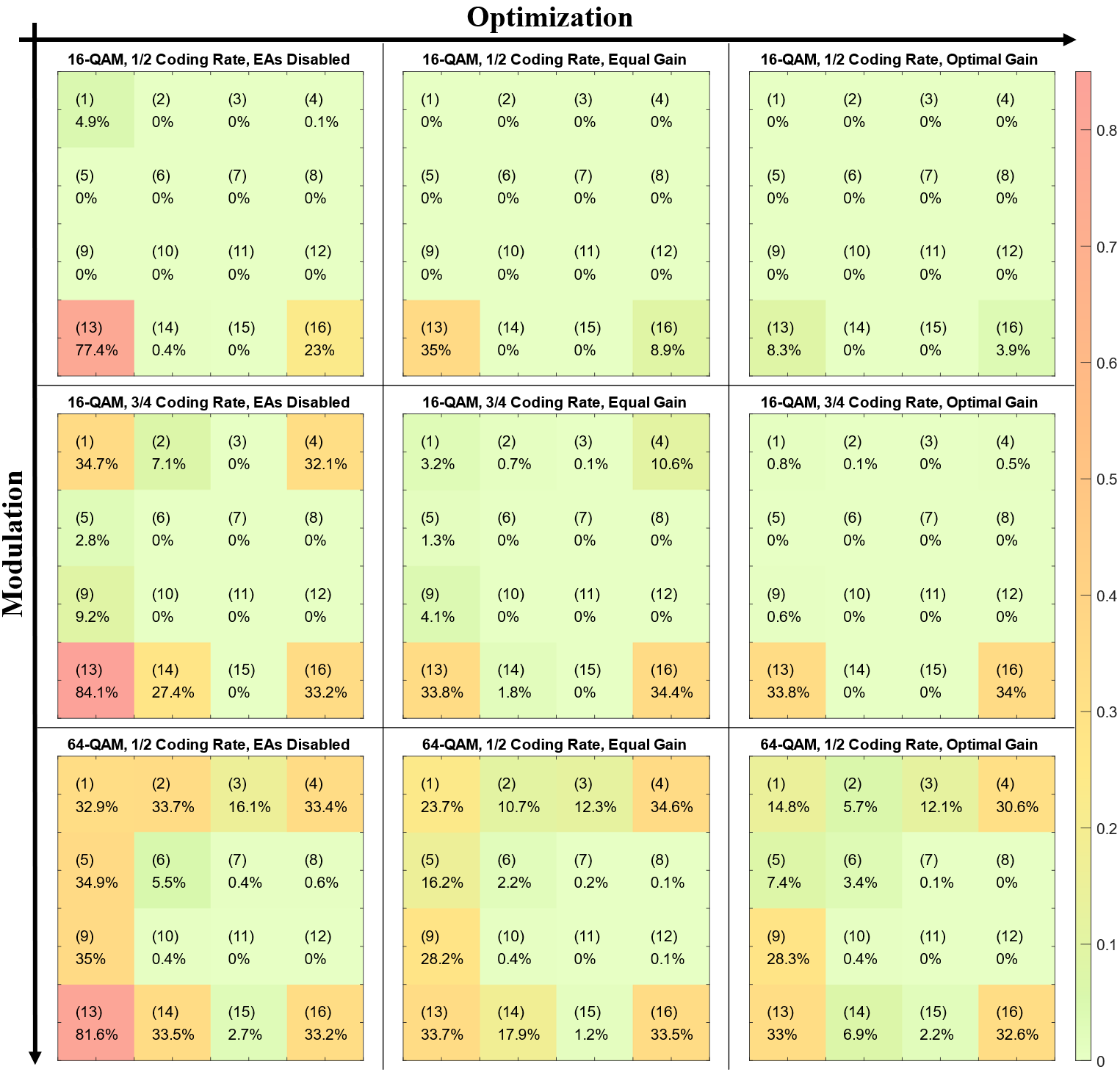}
}
\subfloat[Packet loss rate surf plots.]
{
  \label{plr_stack}
  \includegraphics[width=0.245\textwidth]{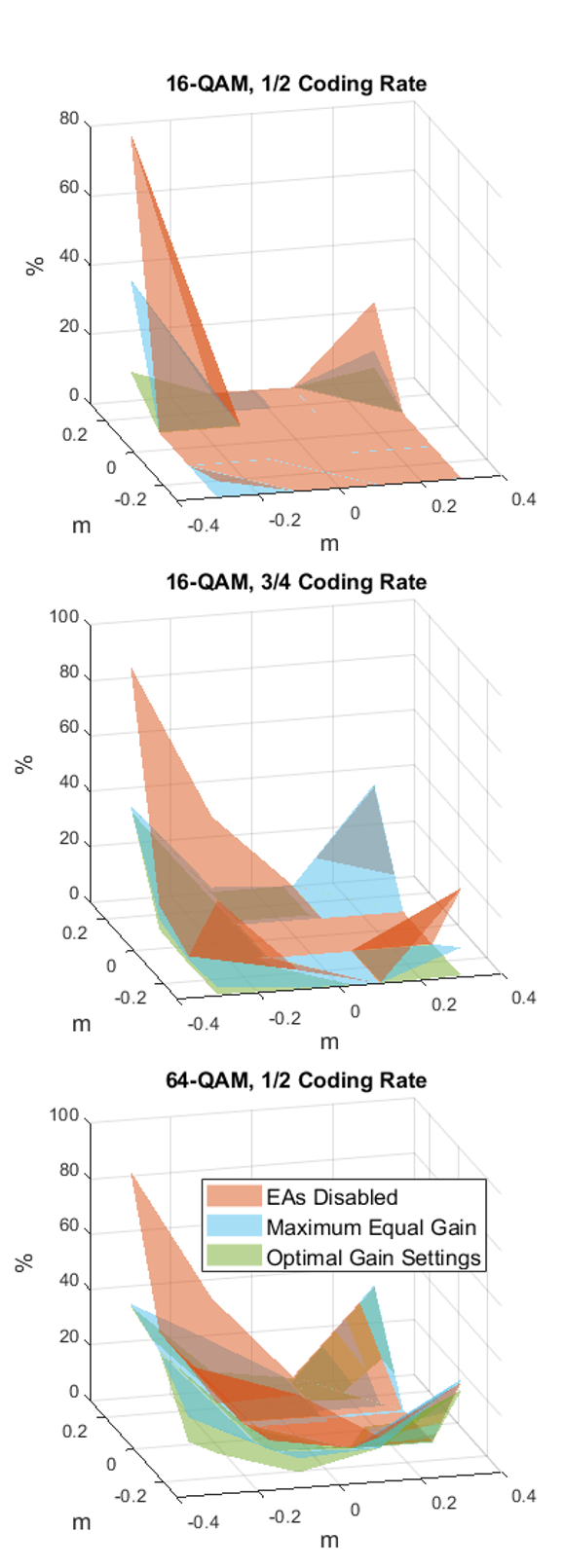}
}
\caption{Packet loss rate measurement results.}
\label{plr_experiments}
\vspace{-0.3cm}
\end{figure*}

\begin{table}
\centering
\begin{tabularx}{0.49\textwidth}{p{1.7cm} p{1.7cm} p{1.7cm} X}
\toprule
 &
  \textbf{\begin{tabular}[c]{@{}l@{}}Disable EAs\end{tabular}} &
  \textbf{\begin{tabular}[c]{@{}l@{}}Maximum\\ Equal Gain\end{tabular}} &
  \textbf{\begin{tabular}[c]{@{}l@{}}Optimal\\ Gain Settings\end{tabular}} \\ \midrule
\textbf{16-QAM, ½} & 6.61 \%  & 2.74 \%  & 0.76 \%  \\
\textbf{16-QAM, ¾} & 14.41 \% & 5.62 \%  & 4.36 \%  \\
\textbf{64-QAM, ½} & 21.50 \% & 13.45 \% & 11.09 \% \\ \bottomrule
\end{tabularx}
\caption{Average packet loss rates over different modulations and EA gain settings.}
\label{average_plr}
\vspace{-0.5cm}
\end{table}
\par
The results are summarized in Fig. \ref{plr_experiments} and Table \ref{average_plr}. In Fig. \ref{plr_experiments}, from left to right, the SNR performance of the deployed EA gain settings improves, while from top to bottom, the channel quality requirement of modulation schemes increases. The subfigure in the top-right corner of Fig. \ref{plr_surf} shows an average packet loss rate of 0.76\%, which is acceptable for most applications. For example, TCP generally works well when the packet loss rate is less than 1\%. Furthermore, Wi-Fi also has a MAC layer ARQ (automatic repeat request) that further decreases the packet loss rate as observed by TCP/IP applications. 
\par
In summary, implementing the optimal EA gain settings significantly improves the packet loss rate, with a maximum reduction of about tenfold, as shown in Table \ref{average_plr}.

\section{Comparison and Connection to Related Work}
The OWE framework pioneers a new approach to achieving ubiquitous optical signal coverage in indoor environments. Its core strength lies in its active, diffuse reflection network, instead of depending on a direct, easily broken path. OWE employs a distributed network of EAs that dynamically reshape the signal's propagation environment. When a dynamic obstruction, such as a moving person or object, blocks a channel, the signal doesn't simply vanish. Instead, it can re-route and propagate through alternative diffuse paths enabled by the distributed EAs (see Appendix C). This offers a significantly higher degree of resilience and adaptability to the constantly changing nature of indoor environments, where blockages are common. OWE's approach effectively creates a "mesh" of optical propagation that is inherently more robust to transient blockages.
\subsubsection{Comparison with Beam-Steering Optical Transceiver Systems}
To extend optical signal coverage, OWE leverages ubiquitous diffuse reflection for seamless connectivity -- unlike systems that integrate optical reconfigurable intelligent surfaces (ORISs) \cite{aboagye2022ris, aboagye2022design} or spatial light modulators (SLMs) \cite{cao2019reconfigurable} into optical transceivers for beam steering, which require precise control and receiver localization \cite{weng2024review}. 
In OWE, EAs are strategically deployed on the ceiling, ensuring proximity to user stations and relaxing the need for precise steering, thus simplifying station hardware design.
\par
Furthermore, stations communicate directly with the nearest overhead EA rather than a distant AP. This proximity reduces transmission power requirements and minimizes the likelihood of signal obstruction by horizontal obstacles such as furniture or moving individuals. These advantages translate to lower power consumption and reduced design complexity for \textbf{mobile devices}, which are crucial for extending battery life and lowering manufacturing costs.
\par
Notably, OWE is not mutually exclusive with existing beam-steering techniques and can be seamlessly integrated with them -- for instance, by mounting ORISs or SLMs on EAs. This potential for hybrid configurations allows for further enhancement of communication quality and optimization of overall system performance.
\subsubsection{Potential for Deploying ORIS with Amplification Ability as EA}
Recent research has introduced an ORIS design that uses reflective surfaces to steer optical signals between transmitters and receivers, combined with liquid crystal technology to amplify the redirected signal \cite{ndjiongue2022design}. However, the achievable gain coefficients in such systems remain limited, typically on the order of tens \cite{ndjiongue2022design, singh2011dependence, aboagye2022design}. This is far below the amplification achievable by OWE’s active analog components, such as cascaded TIAs and PAs in EAs, which can deliver orders of $10^5\times$ amplification in optical signal power.
Due to this limitation, the ORIS cannot compensate for the power loss in OWE’s diffuse reflection channel and thus cannot serve as EAs in the OWE framework.
\par
However, if future advancements significantly improve ORIS amplification capabilities, the theoretical foundations of the OWE framework, including stability constraints and performance optimization, remain applicable to systems employing ORIS-based EAs.
\subsubsection{Latency Benefits Over Conventional Wireless Relays}
Conventional wireless relay networks, whether employing amplify-and-forward or decode-and-forward mechanisms, inherently introduce latency. Amplify-and-forward relays must buffer all or part of an information packet before retransmission to mitigate self-interference \cite{6089444, 6735588}. Decode-and-forward relays incur even greater delays due to additional decoding and signal processing \cite{al2017cognitive, 4273702, al2016decoding}. This latency compromises the user experience for time-sensitive applications and undermines OWC's bandwidth advantages, especially when data packets must traverse multiple "store-and-forward" relay nodes. In contrast, OWE's analog amplification by EAs avoids these buffering and processing delays, preserving OWC's low-latency potential.
\par
\subsubsection{Deployment and Cost-Effectiveness Compared to Multiple APs}
A conventional method for extending wireless coverage is to deploy multiple full-fledged APs. However, each standard AP typically requires sophisticated transceivers, extensive digital processing capabilities, and robust backbone connections for data aggregation. While this approach ensures coverage, it introduces substantial installation costs and operational complexity, particularly in large indoor environments. Moreover, in mobile scenarios, frequent handovers between APs can degrade user experience and introduce additional overhead.
\par
In contrast, OWE provides a more flexible and cost-efficient alternative. Rather than replicating full AP functionalities, OWE extends coverage for a single AP (or a limited number of APs in multi-BSS configurations) through its network of EAs. These EAs are simpler devices dedicated to analog amplification, significantly reducing both deployment costs and complexity. The distributed, modular architecture not only lowers installation overhead but also reduces handovers in mobile scenarios, ensuring a smoother user experience.

\section{Conclusion and Future Work}
This paper introduces OWE, a novel framework that transforms indoor spaces into a dynamically controllable optical propagation medium. OWE's core objective is to overcome the inherent limitations of OWC -- restricted signal coverage due to light's directionality and susceptibility to obstruction -- enabling ubiquitous and reliable optical connectivity.
\par
OWE leverages a distributed network of EAs that act as optical amplifiers, compensating for attenuation while the optical signal propagates through the diffuse reflection channel. By optimizing the programmable gain values across these EAs, OWE can dynamically modify the optical propagation characteristics of the medium to change signal paths, reduce noise, and mitigate interference. 
A key design challenge addressed was preventing amplifier saturation caused by positive feedback loops among the EAs. This was resolved by deriving fundamental stability constraints through systematic analysis and implementing dynamic gain tuning mechanisms.
\par
In simulations, we first demonstrated OWE's ability to significantly extend signal coverage. By deploying optimized gain settings, we further validated OWE's performance: In single-BSS scenarios, optimizing EA gain settings based on station positioning (entry EA selection) remarkably improves SNR compared to a baseline using uniform EA gains without positional consideration. Additionally, in multi-BSS scenarios, OWE ensures fair resource allocation among BSSs while effectively mitigating mutual interference across varying configurations of entry EAs and AP placements.
\par
In experiments, we showed that OWE can significantly reduce packet-loss rates in a Wi-Fi-based TCP/IP network, underscoring its broader applicability and practical benefits.
\par
Future work will focus on developing distributed control schemes for EAs to enable instantaneous, packet-level gain adjustment. Each EA will autonomously optimize its gain to enhance overall OWE network performance, while preserving a fully analog implementation without digital signal processing. By eliminating reliance on a central controller, this method promises greater scalability and adaptability.
\par
In conclusion, OWE lays the groundwork for reimagining wireless environments as programmable optical media. Its capability to extend coverage, optimize channels, and ensure robust, low-latency connectivity positions it as a promising technology for future communication ecosystems.
\appendices
\section{Comprehensive Noise Analysis of the EA}
In the EA circuit, as depicted in Fig. \ref{ea_circuit_model}, the primary noise sources include the \textbf{PD}, \textbf{TIA}, \textbf{PA}, and the \textbf{switching DC-DC converter}. To minimize noise in sensitive analog amplification stages, low-noise linear regulators (LNRs) are used to power the amplifiers, ensuring an ultra-clean supply and negligible power-related noise. However, the LED bias current -- typically large to achieve sufficient optical output -- requires a switching DC-DC converter for higher efficiency and driving capability.
\subsection{PD Noise}
The dominant noise sources in the PD are shot noise and thermal noise \cite{hamamatsuphotonics_2023_si, ghassemlooy2019optical}, whose expressions are given below:
\begin{enumerate}[label=(\roman*)]
\item Shot noise: $\sigma _{\text{shot(PD)}}^2 = 2q\left( {r\left( {{P_r} + {P_{bg}}} \right) + {I_{\text{dark}}}} \right)B$.
\item Thermal noise: $\sigma _{thermal(PD)}^2 = \frac{{4kTB}}{{{R_{sh}}}}$.
\end{enumerate}
The physical constants and parameters in the model are defined as follows: $q$ is the electron charge, $r$ is the photodetector responsivity (A/W), $P_{bg}$ is the background optical power (W), $B$ is the system bandwidth, $k$ is the Boltzmann constant, $T$ is the absolute temperature, and $R_{sh}$ is the shunt resistance of the PD.
\par
The RMS value of the PD noise can be expressed as 
\begin{equation}
{n_{\text{PD}}} = \sqrt {\sigma _{\text{shot(PD)}}^2 + \sigma _{\text{thermal(PD)}}^2} \text{ .}
\end{equation}

\subsection{TIA Noise (Input-Referred)}
The input-referred current noise of the TIA comprises:
\begin{enumerate}[label=(\roman*)]
\item Shot noise arising from the PD current input: \newline $\sigma _{\text{shot(TIA)}}^2 = 2q\left( {r\left( {{P_r} + {P_{bg}}} \right) + {I_{\text{dark}}}} \right)B = \sigma _{\text{shot(PD)}}^2$.
\item Thermal noise from the feedback resistor ${R_{f\left( {\text{TIA}} \right)}}$: \newline $\sigma _{\text{thermal(TIA)}}^2 = \frac{{4kTB}}{{{R_{f\left( {\text{TIA}} \right)}}}}$.
\item Amplifier noise, comprising input-referred voltage noise density (${e_{n(\text{TIA})}}$) and current noise density (${i_{n(\text{TIA})}}$): $\sigma _{amp(\text{TIA})}^2 = \left( {\frac{{e_{n(\text{TIA})}^2}}{{R_{f\left( {\text{TIA}} \right)}^2}} + i_{n(\text{TIA})}^2} \right)B$.
\end{enumerate}
\par
The RMS value of the TIA noise referred to its input is
\begin{equation}
{n_{\text{TIA}}} = \sqrt {\sigma _{\text{shot(TIA)}}^2 + \sigma _{\text{thermal(TIA)}}^2 + \sigma _{\text{amp(TIA)}}^2} \text{ .}
\end{equation}

\subsection{PA Noise (TIA Input-Referred)}
The noise of the PA is modeled as input-referred voltage noise (${e_{n(\text{PA})}}$), current noise (${i_{n(\text{PA})}}$), and resistor thermal noise \cite{vega_2012_source}. These components must be referred to the TIA input to establish a unified noise model.
\par
For MOSFET-input PAs, the current noise density $i_{n(\text{PA})}$ often negligible. In an inverting amplifier configuration with voltage gain ${G_{\text{PA}}} =  - \frac{{{R_{f\left( {\text{PA}} \right)}}}}{{{R_{i\left( {\text{PA}} \right)}}}}$, the signal gain magnitude is $\frac{{{R_{f\left( {\text{PA}} \right)}}}}{{{R_{i\left( {\text{PA}} \right)}}}}$. However, the noise gain magnitude is $1 + \frac{{{R_{f\left( {\text{PA}} \right)}}}}{{{R_{i\left( {\text{PA}} \right)}}}}$ because the noise sees the non-inverting path. The signal gain is used for referring noise to the input, while the noise gain determines the amplification of ${e_{n(\text{PA})}}$ and ${i_{n(\text{PA})}}$ \cite{karki_2003_calculating}.
\begin{enumerate}[label=(\roman*)]
\item The thermal noise from ${R_{i\left( {\text{PA}} \right)}}$ is inherently input-referred, while the thermal noise from ${R_{f\left( {\text{PA}} \right)}}$ is divided by the signal gain when referred to the input. The total PA resistor thermal noise referred to the TIA input is given by
\begin{align} 
&\sigma _{\text{thermal(PA,TIA)}}^2 \notag
\\ &= \frac{{4kTB{R_{f\left( {\text{PA}} \right)}}}}{{{{\left( {\frac{{{R_{f\left( {\text{PA}} \right)}}}}{{{R_{i\left( {\text{PA}} \right)}}}}} \right)}^2}{{\left| {{Z_{\text{TIA}}}} \right|}^2}}} + \frac{{4kTB{R_{i\left( {\text{PA}} \right)}}}}{{{{\left| {{Z_{\text{TIA}}}} \right|}^2}}} \notag
\\ & = \frac{{4kTB{R_{f\left( {\text{PA}} \right)}}}}{{G_{\text{PA}}^2{{\left| {{Z_{\text{TIA}}}} \right|}^2}}} + \frac{{4kTB{R_{i\left( {\text{PA}} \right)}}}}{{{{\left| {{Z_{\text{TIA}}}} \right|}^2}}} \text{ .}
\end{align}
\item The voltage noise ${e_{n(\text{PA})}}$ is multiplied by the noise gain at the output, but when referred to the input, it is divided by the signal gain. The PA voltage noise referred to the TIA input is given by
\begin{align}
& \sigma _{\text{voltage(PA,TIA)}}^2 \notag
\\ & = \frac{{e_{n(\text{PA})}^2}}{{{{\left| {{Z_{\text{TIA}}}} \right|}^2}}}{\left( {\frac{{1 + \frac{{{R_{f\left( {\text{PA}} \right)}}}}{{{R_{i\left( {\text{PA}} \right)}}}}}}{{\frac{{{R_{f\left( {\text{PA}} \right)}}}}{{{R_{i\left( {\text{PA}} \right)}}}}}}} \right)^2}B \notag
\\ &= \frac{{e_{n(\text{PA})}^2}}{{{{\left| {{Z_{\text{TIA}}}} \right|}^2}}}{\left( {\frac{{1 + {G_{\text{PA}}}}}{{{G_{\text{PA}}}}}} \right)^2}B \text{ .}
\end{align}
\item The current noise ${i_{n(\text{PA})}}$ contribution depends on ${R_{f\left( {\text{PA}} \right)}}\parallel {R_{i\left( {\text{PA}} \right)}}$. The PA current noise referred to the TIA input is given by 
\begin{align}
&\sigma _{\text{current(PA,TIA)}}^2 \notag
\\ & = \frac{{i_{n(\text{PA})}^2{{\left( {{R_{f\left( {\text{PA}} \right)}}\parallel {R_{i\left( {\text{PA}} \right)}}} \right)}^2}}}{{{{\left| {{Z_{\text{TIA}}}} \right|}^2}}}{\left( {\frac{{1 + \frac{{{R_{f\left( {\text{PA}} \right)}}}}{{{R_{i\left( {\text{PA}} \right)}}}}}}{{\frac{{{R_{f\left( {\text{PA}} \right)}}}}{{{R_{i\left( {\text{PA}} \right)}}}}}}} \right)^2}B \notag
\\ & = \frac{{i_{n(\text{PA})}^2{{\left( {{R_{f\left( {\text{PA}} \right)}}\parallel {R_{i\left( {\text{PA}} \right)}}} \right)}^2}}}{{{{\left| {{Z_{\text{TIA}}}} \right|}^2}}}{\left( {\frac{{1 + {G_{\text{PA}}}}}{{{G_{\text{PA}}}}}} \right)^2}B \text{ .}
\end{align}
\end{enumerate}
In the noise models above, ${Z_{\text{TIA}}}$ denotes the impedance of the TIA. We assumes ${Z_{\text{TIA}}} = {R_{f\left( {\text{TIA}} \right)}}$ for simple resistive feedback in the TIA.
\par
The RMS value of the PA noise referred to the input of the TIA is
\begin{equation}
{n_{\text{PA,TIA}}} = \sqrt {\sigma _{\text{thermal(PA,TIA)}}^2 + \sigma _{\text{voltage(PA,TIA)}}^2 + \sigma _{\text{current(PA,TIA)}}^2} \text{ .}
\end{equation}

\subsection{DC-DC Converter Noise}
The power spectral density (PSD) of noise introduced by the DC-DC switching converter consists of two distinct components:
\begin{enumerate}[label=(\roman*)]
\item Switching noise: Characterized by spectral peaks at the switching frequency and its harmonics.
\item Broadband noise: Exhibiting a flat noise floor across the frequency spectrum.
\end{enumerate}
\par
The switching noise is modeled as Gaussian spectral peaks centered at the fundamental switching frequency $f_{sw}$ and its harmonics $kf_{sw}$, where $k=1,2,3$, corresponding to the fundamental frequency and the first two harmonic components. The PSD of the switching noise (voltage noise) is given by 
\begin{equation}
{S_{\text{V,switching}}}\left( f \right) = A\sum\limits_{k = 1}^3 {\exp \left( { - \frac{{\left( {f - k{f_{sw}}} \right)^2}}{{2{{\left( {0.1{f_{sw}}} \right)}^2}}}} \right)} \text{ ,}
\end{equation}
where $A = {10^{ - 9}}{\text{ V}^2}\text{/Hz}$ is the scaling factor for noise power. $0.1{f_{sw}}$ determines the bandwidth of each peak, that is the standard deviation of the Gaussian.
\par
The broadband noise represents thermal noise and other random fluctuations that are present across all frequencies. We model it as a constant noise floor, denoted by $b = {10^{ - 12}} {\text{ V}^2}\text{/Hz}$. This white noise component establishes a baseline noise level and is frequency-independent.
\par
The PSD of DC-DC converter's output voltage noise is given by
\begin{equation}
{S_{\text{V(DC-DC)}}}\left( f \right) = {S_{\text{V,switching}}}\left( f \right) + b \text{ .}
\end{equation}
\subsubsection{Bias-Tee Low-Pass Filtering}
The bias-tee consists of an inductor $L$ and forms a low-pass filter with the LED’s current-limiting resistor $R_{\text{BT}}$. The cutoff frequency $f_c$ is
\begin{equation}
{f_c} = \frac{{{R_{\text{BT}}}}}{{2\pi L}} \text{ .}
\end{equation}
The noise of the DC-DC converter is attenuated above the cutoff frequency $f_c$. The transfer function of the low-pass filter is
\begin{equation}
{H_{\text{BT}}}\left( f \right) = \frac{1}{{1 + j\frac{f}{{{f_c}}}}} \text{ .}
\end{equation}
The magnitude of the transfer function is
\begin{equation}
\left| {{H_{\text{BT}}}\left( f \right)} \right| = \frac{1}{{\sqrt {1 + {{\left( {\frac{f}{{{f_c}}}} \right)}^2}} }} \text{ .}
\end{equation}
\subsubsection{Noise Current Calculation}
The current noise contributed by the DC-DC converter to the LED bias current is characterized by
\begin{equation}
{S_{I\left( \text{DC-DC} \right)}}\left( f \right) = \frac{{{S_{V(\text{DC-DC})}}\left( f \right)}}{{{Z_{\text{LED}}}\left( f \right)}}{\left| {{H_{\text{BT}}}\left( f \right)} \right|^2} \text{ ,}
\end{equation}
where $Z_{\text{LED}}(f)$ represents the impedance of the LED branch. For simplicity, we can approximate ${Z_{\text{LED}}}\left( f \right) \approx {R_{\text{BT}}}$, since the LED’s small-signal impedance at the DC-biased operating point (set by the bias tee) is negligible.
\par
The total current noise contributed by the DC-DC converter equals the integral of its power spectral density ${S_{I\left( {\text{DC-DC}} \right)}}\left( f \right)$ over the system bandwidth:
\begin{align}
&{n_{\text{DC-DC}}} = \sqrt {\int\limits_0^B {{S_{I\left( {\text{DC-DC}} \right)}}\left( f \right)df} } \notag
\\ & \approx \sqrt {\int\limits_0^B {\left[ {A\sum\limits_{k = 1}^3 {\exp \left( { - \frac{{\left( {f - k{f_{sw}}} \right)^2}}{{2{{\left( {0.1{f_{sw}}} \right)}^2}}}} \right)}  + b} \right] \cdot \frac{{{{\left| {{H_{\text{BT}}}\left( f \right)} \right|}^2}}}{{{R_{\text{BT}}}}}df}} \text{ .} 
\end{align}
\subsection{EA Signal Flow}
\subsubsection{Output of PA}
The TIA input current from the PD consists of two components: ${s_{\text{PD}}} + {n_{\text{PD}}}$, where ${s_{\text{PD}}} = r{P_r}$ is the signal current and $n_{\text{PD}}$ is the noise current. Recall that we previously calculated all noise contributions referred to the TIA input. Thus, the equivalent input signal at the TIA input becomes ${s_{\text{PD}}} + {n_{\text{PD}}} + {n_{\text{TIA}}} + {n_{\text{PA,TIA}}}$. Consequently, the output voltage after amplification through both the TIA and PA stages is given by
\begin{equation}
{V_{\text{PA}}} = {Z_{\text{TIA}}}{G_{\text{PA}}}\left( {{s_{\text{PD}}} + {n_{\text{PD}}} + {n_{\text{TIA}}} + {n_{\text{PA,TIA}}}} \right) \text{ .}
\end{equation}
\subsubsection{Output Optical Power of EA}
The PA output drives the LED through a series resistor $R_{\text{BT}}$, which limits the current to prevent excessive loading on the amplifier and avoid overdriving the LED. Due to the LED's negligible small-signal impedance at its DC operating point, $R_{\text{BT}}$ dominates the PA's effective load impedance. Therefore, the LED current can be approximated as
\begin{equation}
{I_{\text{LED}}} = \frac{{{V_{\text{PA}}}}}{{{R_{\text{BT}}}}}+ {n_{\text{DC-DC}}} \text{ .}
\end{equation}
The optical power $P_t$ (in Watt) emitted by an LED is approximately proportional to the forward current ${I_{\text{LED}}}$ (in Ampere) within the low-to-moderate current range. A simplified linear model for the output optical power as a function of current is given by
\begin{equation}
{P_t} = \eta {U_{\text{LED}}}{I_{\text{LED}}} \text{ .}
\end{equation}
where $\eta$ is LED radiant efficiency, defined as the ratio of emitted visible optical power to input electrical power. $U_{\text{LED}}$ denotes the LED forward voltage, which is the voltage drop across the device when operating at its rated current.
\subsubsection{EA Output SNR}
The LED current comprises both signal and noise components:
\begin{equation}
{I_{\text{LED}}} = {s_{\text{LED}}} + {n_{\text{LED}}} \text{ ,}
\end{equation}
where the signal current through the LED is given by ${s_{\text{LED}}} = \frac{{{s_{\text{PD}}}{Z_{\text{TIA}}}{G_{\text{PA}}}}}{{{R_{\text{BT}}}}}$.
The corresponding LED noise current can be expressed as
\begin{equation}
{n_{\text{LED}}} = \frac{{{G_{\text{PA}}}{Z_{\text{TIA}}}}}{{{R_{\text{BT}}}}}\left( {{n_{\text{PD}}} + {n_{\text{TIA}}} + {n_{\text{PA,TIA}}}} \right) + {n_{\text{DC-DC}}} \text{ .}
\end{equation}
The SNR of the electrical signal output of EA driving the LED can be written as
\begin{equation}
\text{SNR}_{\text{EA}} = \frac{{s_{\text{LED}}^2}}{{n_{\text{LED}}^2}} \text{ .}
\end{equation}
Since the optical power is proportional to the LED current, the optical SNR equals the square root of the electrical SNR ($\text{SNR}_{\text{EA}}$). However, because both the AP and station perform signal sampling and processing in the electrical domain, we use electrical SNR as the performance metric for our system.

\section{Explanation of Simulation Parameter Settings}
The half-power angle of the optical transmitter and the acceptance angle of the optical receiver are calculated based on illumination requirements and signal coverage needs.
\subsection{Half Power Angle Calculation}
\begin{figure}
\centering
\includegraphics[width=0.5\textwidth]{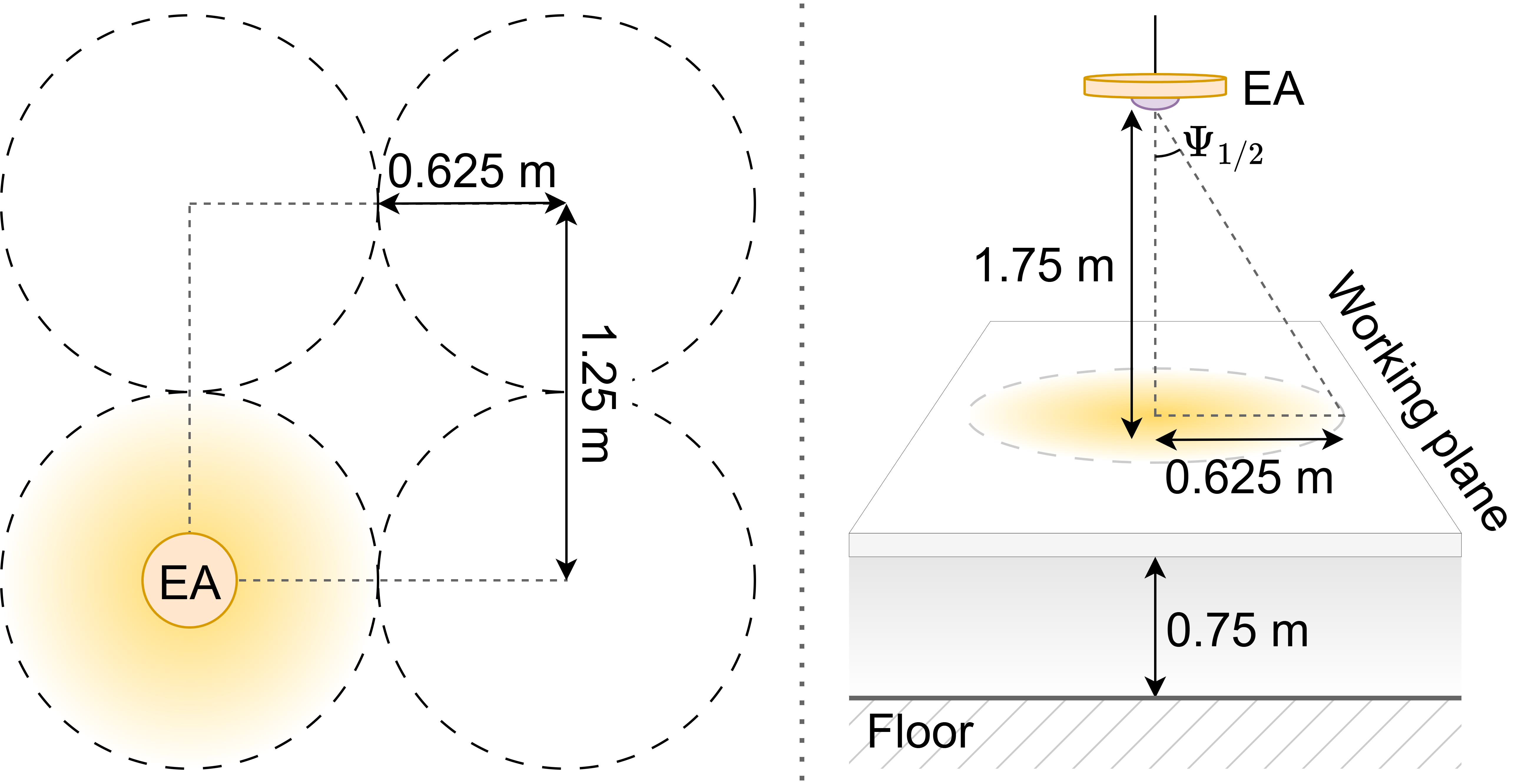}
\caption{Half-power angle configuration based on illumination uniformity requirements.}
\label{half_power_angle}
\vspace{-0.1cm}
\end{figure}

The half-power angle of a luminaire is the angular range over which the emitted optical power (or luminous intensity) exceeds $50\%$ of its peak value, defining the coverage area. To ensure uniform illuminance, the projected circular areas of adjacent luminaires -- determined by their half-power angles -- should at least be tangent to one another on the working plane (typically positioned $0.75 \text{\ m}$ above the floor). This arrangement eliminates dark spots in the illuminated area, as illustrated in Fig. \ref{half_power_angle}.
By substituting numerical values into the model, the half-power angle of the optical transmitter in the EA can be calculated as follows:
\begin{equation}
{\Psi _{1/2}} = {\rm{atan}}\left( {\frac{{0.625}}{{1.75}}} \right) = {\rm{19}}{\rm{.65}}^\circ \text{ .}
\end{equation}

\subsection{Acceptance Angle Calculation}
\begin{figure}
\centering
\includegraphics[width=0.5\textwidth]{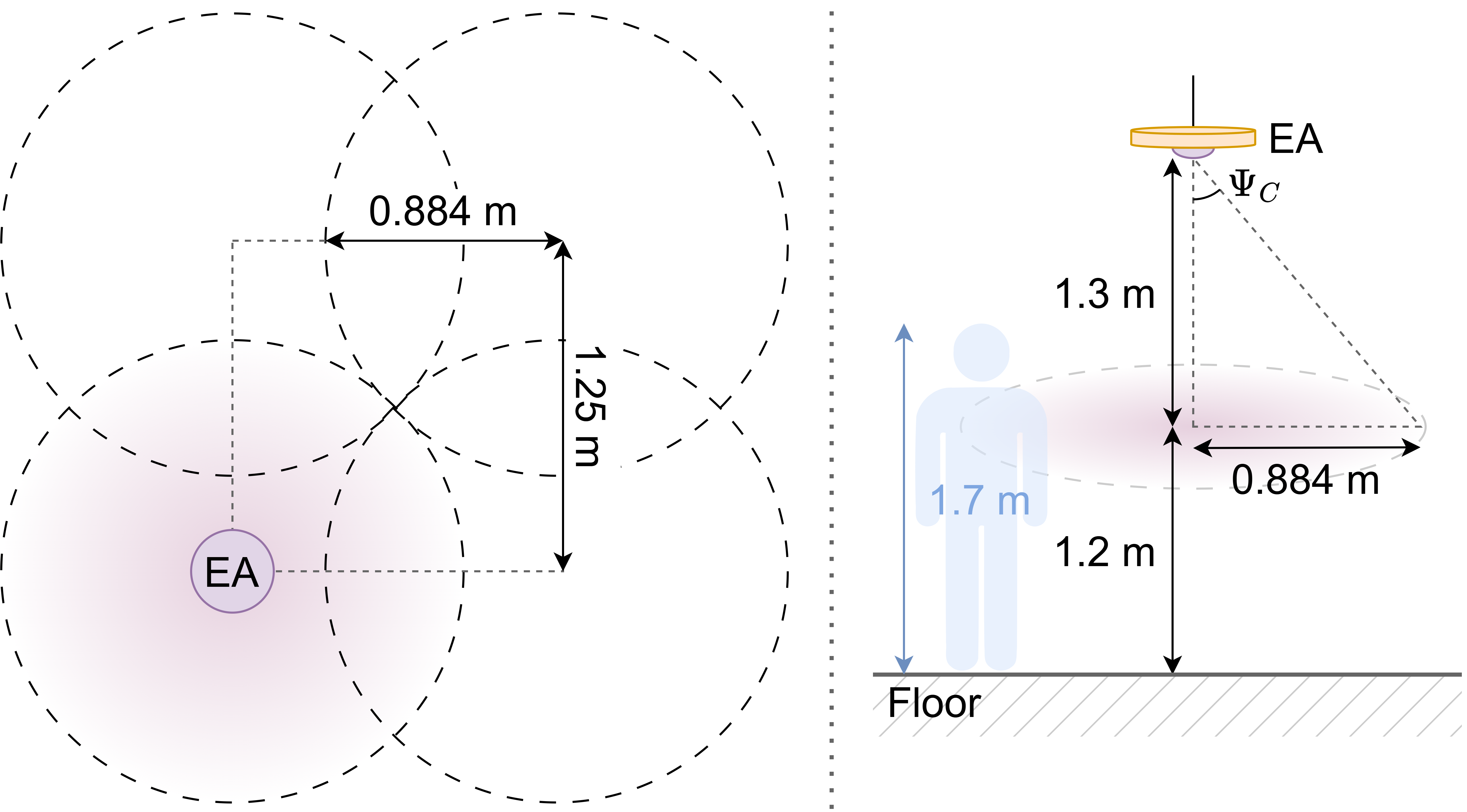}
\caption{Acceptance angle configuration based on signal coverage requirements.}
\label{acceptance_angle}
\vspace{-0.5cm}
\end{figure}

To ensure seamless signal reception without blind spots, the receiver's acceptance angle must provide full coverage across all possible user device locations on the horizontal plane. Since smartphone heights typically range from approximately $0.75 \text{ m}$ (when placed on a table) to $0.9 \text{ - } 1.2 \text{ m}$ (when held for typing or viewing, accounting for elbow rest height), the acceptance angle must accommodate this variation. The relationship between the signal reception area and the distance from the EA to the user's elbow (corresponding to a typical device-holding height of $1.2 \text{ m}$) is illustrated in Fig. \ref{acceptance_angle}.
By substituting numerical values into the model, the acceptance angle of the optical receiver in the EA can be calculated as follows:
\begin{equation}
{\Psi _C} = {\rm{atan}}\left( {\frac{{0.884}}{{1.3}}} \right) = {\rm{34}}{\rm{.21}}^\circ \text{ .}
\end{equation}

\section{Dynamic Gain Optimization in OWE}
OWE dynamically optimizes the signal's propagation environment by intelligently coordinating gain settings across distributed EAs. When a moving obstruction, such as a person or object, blocks the original signal path, the system does not simply lose connectivity. Instead, it leverages alternative paths enabled by the distributed EAs, ensuring seamless signal rerouting and rapid connection recovery.
\par
For the OWE system that uses a central controller to optimize EA gain settings, two key pieces of information are required: 1) Localization of the user station (identifying the entry EA receiving the signal), and 2) Diffuse channel responses between EAs.

\subsection{User Station Localization}
The positions of user stations can be determined through:
\begin{itemize}
    \item Vision-based approaches (e.g., ceiling-mounted cameras),
    \item Optical-based sensing techniques (e.g., tracking changes in channel state information (CSI) or received signal strength (RSS) among EAs).
\end{itemize}
Notably, meter-level localization accuracy suffices for selecting the appropriate entry EA, eliminating the need for highly precise positioning.

\subsection{Channel Response Estimation Between EAs}
The strategy for measuring inter-EA channel responses depends on environmental dynamics:

\begin{figure*}
\centering
\includegraphics[width=1\textwidth]{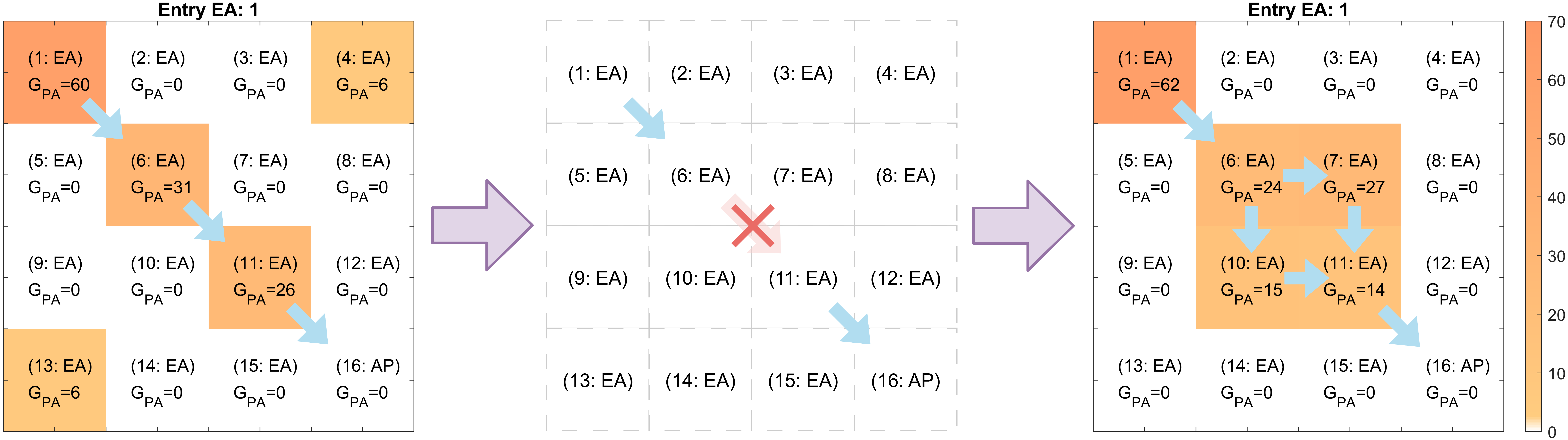}
\caption{Simulation of OWE bypassing a blocked optical channel (EA6 $\rightarrow$ EA11).}
\label{dynamic_control}
\vspace{-0.3cm}
\end{figure*}

\begin{itemize}
    \item Static Environments (e.g., smart factories): Channel responses can be pre-measured during deployment and treated as constants. This configuration is suitable for controlled settings such as automated warehouses or robotic assembly lines where environmental changes are minimal.
    \item Dynamic Environments (e.g., offices or public spaces): If a diffuse reflection channel between EAs is blocked (e.g., by moving objects), the system employs an adaptive control mechanism to mitigate the effects. The controller first identifies blockages by actively probing channel responses along the signal path. Upon detecting an obstruction, it performs two key actions: 1) Updating the channel response knowledge, 2) Optimizing EA gain settings accordingly to reroute signals through alternative paths to maintain connectivity, as demonstrated in Fig. \ref{dynamic_control}. The technical implementation of this channel probing and estimation process will be detailed in the following section.
\end{itemize}
\begin{figure}
\centering
\includegraphics[width=0.35\textwidth]{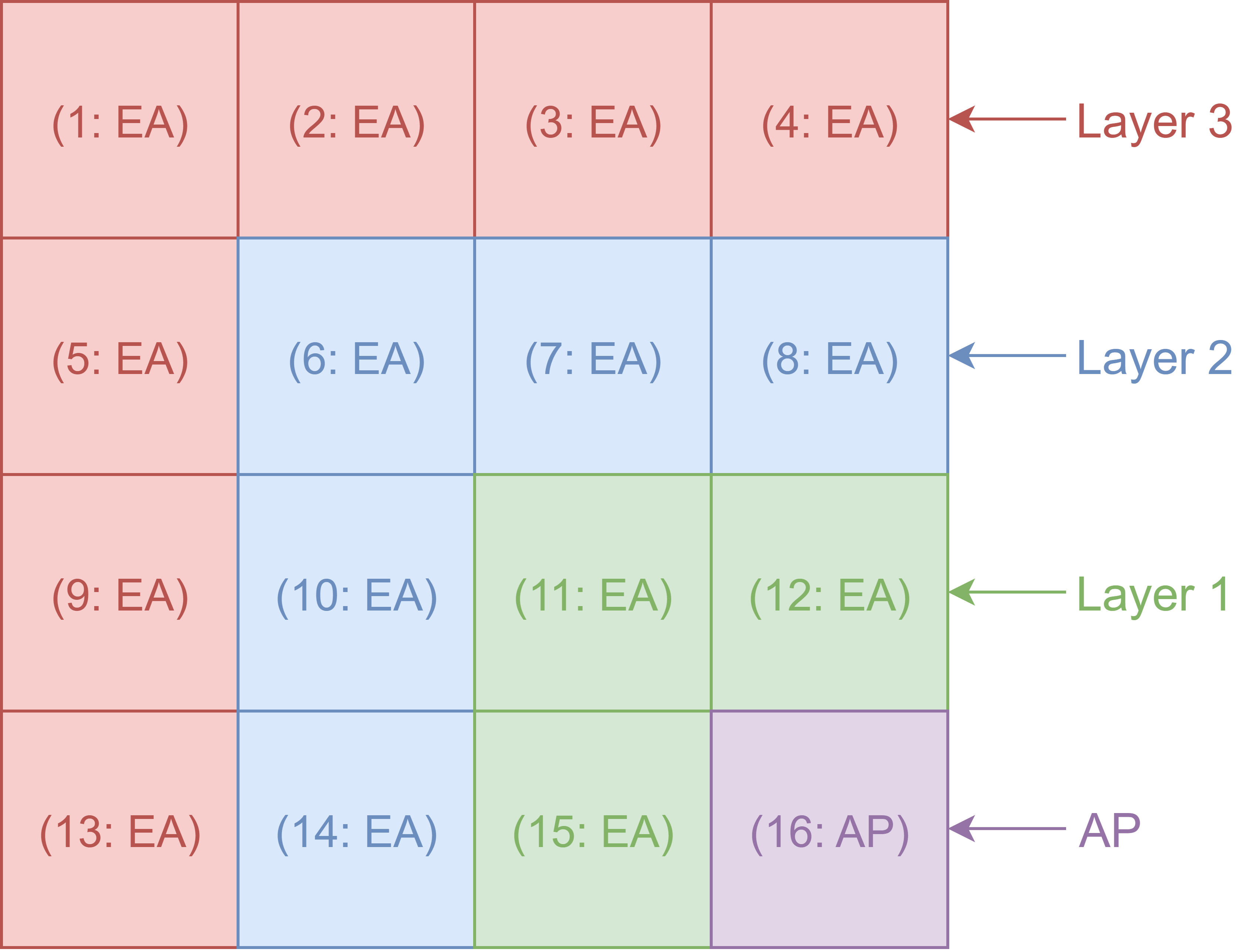}
\caption{Layered grouping of EAs in the OWE based on hop distance from the AP.}
\label{ea_layers}
\vspace{-0.4cm}
\end{figure}
\subsection{Automated Channel Estimation Mechanism}
To adapt to dynamic environments, we propose a channel estimation protocol for measuring the channel responses between EAs. Each EA is equipped with a simple tone generator to emit a known optical reference signal. The protocol operates as follows:
\begin{enumerate}
    \item \textbf{Blockage Detection Between EAs}:
    In dynamic environments, moving obstacles can obstruct signal propagation paths. To maintain optimal performance, the OWE system should detect such blockages, update the channel response information, and re-optimize the EA gain settings accordingly.
    \begin{itemize}
        \item Blockage detection (e.g., between EA6 and EA11 in Fig. \ref{dynamic_control}) involves only the EAs along the target signal path (e.g., EA1 $\rightarrow$ AP). All other EAs retain their original gain settings, ensuring isolation between the measured path and operational paths.
        \item The detection process involves sequential testing
        \begin{itemize}  
            \item Only one EA along the measured path transmits the reference signal at a time, with its gain set to 0, while all others retain their pre-blockage gain values.
            \item Testing begins with the EA closest to the AP (e.g., EA11). If the AP successfully receives the signal, the path is confirmed as unobstructed.  
            \item The next EA in the path (e.g., EA6) then emits the signal. If the AP fails to detect it, the blockage is localized between the two EAs.  
        \end{itemize}
        This stepwise approach ensures precise identification of the obstruction location.  
    \end{itemize}

    \item \textbf{Channel Response Probing from Scratch}: 
    During the initial deployment phase or when the network structure changes (e.g., when new EAs are added), the channel response information may need to be collected from scratch or refreshed. This process can be automated, eliminating the need for manual measurements.
    \begin{itemize}
        \item Layer-based EA Activation
        \begin{itemize}
            \item EAs are grouped into "layers" based on their hop distance from the AP, as depicted in Fig. \ref{ea_layers}.
            \item Layer 1: EAs directly adjacent to the AP (1-hop).
            \item Layer 2: EAs requiring one intermediate EA (2-hop), etc.
        \end{itemize}
        
        \item During measurement, only one EA in the target layer transmits the tone, while all other EAs in the same layer, along with those \textbf{not} involved in signal propagation in the inner and outer layers, are muted (gain = 0) to isolate the measured path. Meanwhile, the EAs in the inner layers responsible for propagating the tone to the AP are set to a known gain value.

        \item Response Inference
        \begin{itemize}
            \item First Layer Estimation: The channel responses between the AP and its directly adjacent EAs are derived by comparing the received signal with the reference tone.
            \item Higher-Layer Estimation (e.g., Layer 2): The controller utilizes the known gains of Layer 1 EAs along with pre-measured Layer 1 channel responses to infer the channel responses for Layer 2.
        \end{itemize}
        This process is extended iteratively, leveraging known EA gains and inner-layer channel responses to deduce the responses for subsequent outer layers.
    \end{itemize}
    
    \item \textbf{Potential Optimization of Measurement Overhead}:
    To enhance efficiency, the measurement process can be optimized in the following ways:
    \begin{itemize}
        \item Priority-based Measurement Scheduling: Measurements are prioritized for EAs along active or critical paths to minimize latency.
        \item Redundancy Reduction: Redundant measurements (e.g., between rarely used EAs) can be skipped to reduce overhead.
        \item Proactive Channel Prediction: By proactively estimating channel conditions between EAs, signals can be rerouted preemptively, ensuring seamless service continuity.
    \end{itemize}
\end{enumerate}

\bibliography{references}

\end{document}